\documentclass[11pt,notitlepage,eqsecnum,amsmath,preprintnumbers,superscriptaddress,nofootinbib,aps]{revtex4-1} 
\pdfoutput=1
\usepackage{graphicx}
\usepackage{subfigure}
\usepackage{hyperref}
\usepackage{array}
\usepackage{amsmath}
\usepackage{graphicx}
\usepackage{amssymb}
\usepackage{subfigure}
\usepackage{epsfig}
\usepackage{color}

\usepackage{multirow}

\usepackage{booktabs}
\usepackage[x11names,dvipsnames,table]{xcolor} 
\usepackage{colortbl}

\usepackage{xcolor}
\usepackage{colortbl}

\definecolor{Gray}{gray}{0.85}
\definecolor{LightGray}{gray}{0.93}
\definecolor{LightGreen}{rgb}{0.88, 1, 0.88}
\definecolor{LightCyan}{rgb}{0.88,1,1}
\definecolor{LightRed}{rgb}{1, 0.85, 0.85}
\definecolor{LightRed}{rgb}{1, 0.88, 0.88}
\definecolor{LightYellow}{rgb}{1, 1, 0.85}
\definecolor{LightBlue}{rgb}{0.87, 0.94, 1}
\definecolor{white}{gray}{1}

\newcolumntype{G}{>{\columncolor{LightGray}}c}

\long\def\del #1 \enddel { }

\usepackage{graphicx}

\usepackage{subfigure}

\usepackage{hyperref}

\def\Rds{R_{{}_{\rm dS}}}
\def\RLB{R_{{}_{\rm LB}}}
\def\RLB{R_{L}}
\def\beq{\begin{equation}}
\def\eeq{\end{equation}}

\def\bea{\arraycolsep .1em \begin{eqnarray}}
\def\eea{\end{eqnarray}}
\def\Tr{{\rm Tr}}

\def\eq#1{(\ref{#1})}

\def\R{R}
\def\s0#1#2{\mbox{\small{$ \frac{#1}{#2} $}}}
\def\0#1#2{\frac{#1}{#2}}

\def\grgl{\:\hbox to -0.2pt{\lower2.5pt\hbox{$\sim$}\hss}{\raise3pt\hbox{$>$}}\:}
\def\klgl{\:\hbox to -0.2pt{\lower2.5pt\hbox{$\sim$}\hss}{\raise3pt\hbox{$<$}}\:}
\def \lta {\mathrel{\vcenter
     {\hbox{$<$}\nointerlineskip\hbox{$\sim$}}}}

\makeatletter
    \def\CT@@do@color{%
      \global\let\CT@do@color\relax
            \@tempdima\wd\z@
            \advance\@tempdima\@tempdimb
            \advance\@tempdima\@tempdimc
    \advance\@tempdimb\tabcolsep
    \advance\@tempdimc\tabcolsep
    \advance\@tempdima2\tabcolsep
            \kern-\@tempdimb
            \leaders\vrule
                    \hskip\@tempdima\@plus  1fill
            \kern-\@tempdimc
            \hskip-\wd\z@ \@plus -1fill }

   \makeatother
   
\begin{document}

\title{On de Sitter solutions in asymptotically safe $f(R)$ theories}
\author{K.~Falls}
\affiliation{Institut f\"{u}r Theoretische Physik, University of Heidelberg, 69120 Heidelberg, Germany}
\author{D.F.~Litim}
\author{K.~Nikolakopoulos}
\affiliation{Department of Physics and Astronomy, University of Sussex, Brighton, BN1 9QH, U.K.}
\author{C.~Rahmede}
\affiliation{Rome International Centre for Material Science Superstripes RICMASS, via dei Sabelli 119A,
00185 Roma, Italy}

\begin{abstract}
The availability of scaling solutions in renormalisation group improved versions of cosmology are investigated in the high-energy limit. We adopt $f(R)$-type models of quantum gravity which display an interacting ultraviolet fixed point at shortest distances.
Expanding the gravitational fixed point action to very high order in the curvature scalar, we detect a convergence-limiting singularity in the complex field plane. Resummation techniques including Pad\'e approximants as well as infinite order approximations of the effective action are used to maximise the domain of validity. We  find that the theory displays near de Sitter  solutions as well as an anti-de Sitter solution in the UV whereas real de Sitter solutions, for small curvature,  appear to be absent. The significance of our results for inflation, and implications for more general models of quantum gravity are discussed.   
\end{abstract}

\maketitle
\tableofcontents

\newpage
\section{\bf Introduction}

The asymptotic safety conjecture for gravity  \cite{Weinberg:1980gg} stipulates that a quantum theory of  gravity may very well exist as a conventional local quantum field theory provided it develops an interacting UV fixed point  at highest energies. 
In recent years growing  evidence for asymptotic safety has been accumulated, largely based on increasingly sophisticated renormalisation group studies of gravity without \cite{Reuter:1996cp,Dou:1997fg,Souma:1999at,
Lauscher:2001ya,Lauscher:2002sq,Litim:2003vp,Fischer:2006fz,Fischer:2006at,Litim:2006dx,Codello:2006in,
Codello:2007bd,Machado:2007ea,Codello:2008vh,Litim:2008tt,
Benedetti:2012dx,Demmel:2012ub,Christiansen:2012rx,Dietz:2012ic,
Falls:2013bv,
Falls:2014tra,Christiansen:2014raa,Saltas:2014cta,
Eichhorn:2015bna,Demmel:2015oqa,Falls:2015qga,Ohta:2015efa}  
or with matter fields 
\cite{Percacci:2002ie,Percacci:2003jz,
Narain:2009fy,Narain:2009gb,Eichhorn:2011pc,Eichhorn:2012va,Folkerts:2011jz,Harst:2011zx,Zanusso:2009bs,
Eichhorn:2009ah,Eichhorn:2010tb,Groh:2010ta,Meibohm:2015twa}  (see \cite{Litim:2011cp} for a  review), including the recent  proof of existence for  asymptotic safety  in $4d$ quantum gauge theories \cite{Litim:2014uca,Bond:2016dvk,Bond:2017wut,Bond:2017lnq,Bond:2017suy,Bond:2017tbw,Bond:2018oco}, without gravity.  In general the fluctuations of the metric field are found to be strong, strong  enough for gravity to become anti-screening such that Newton's coupling weakens quantum-mechanically towards shorter distances \cite{Litim:2006dx,Niedermaier:2006ns,Niedermaier:2006wt}.  On the other hand, it has also been observed that higher order gravitational interactions lead to near-Gaussian scaling \cite{Falls:2013bv}, suggesting that quantum effects for these are somewhat less pronounced. It would thus seem that quantum gravity becomes ``as Gaussian as it gets'' \cite{Falls:2014tra} despite of its perturbative non-renormalisability \cite{Gies:2016con}. In the language of critical phenomena, gravitational couplings invariably display interacting fixed points modifying the short distance behaviour of the theory with a critical surface of low dimensionality and characteristic non-classical scaling exponents for a few relevant couplings including Newton's coupling and the cosmological constant,  together with near Gaussian scaling for its irrelevant higher order couplings. 

Cosmology, thanks to the wealth of data available from observation \cite{Ade:2013uln,Ade:2013zuv,Perlmutter:1998np}, offers an important territory to test the asymptotic safety scenario for gravity. Provided that asymptotic safety is realised in nature, it is conceivable that the characteristic quantum gravitational modifications  have impacted during the very early universe,  including its phase of inflationary expansion and the phase of  late time acceleration. A number of studies have explored  these possibilities 
by exploiting characteristics of an asymptotically safe fixed point using renormalisation group improvements of the effective action or of the gravitational equations of motion including those of Friedmann-Robertson-Walker universes \cite{Shapiro:2000dz,Weinberg:2009wa,Reuter:2005kb,Bonanno:2001xi,Bonanno:2001hi,
Bonanno:2009nj,Bonanno:2010bt,Bonanno:2012jy,Frolov:2011ys,Cai:2011kd,
Hindmarsh:2012rc,Koch:2010nn,Copeland:2013vva,Bonanno:2015fga,
Contillo:2011ag,Hindmarsh:2011hx,Ahn:2011qt,
Kaya:2013bga,Saltas:2015vsc, Kofinas:2016lcz}.

In this paper, we are particularly interested in de Sitter solutions for cosmology, and whether these may arise through fluctuations of the metric field. To answer this question, the quantum effective action in the fixed point regime 
needs to be available. Here, we will  exploit polynomial approximations of the gravitational action up to very high order in the Ricci scalar curvature which have been made available in \cite{Falls:2013bv,Falls:2014tra}. Our study extends earlier investigations  \cite{Codello:2007bd,Codello:2008vh,Machado:2007ea,Bonanno:2010bt} to substantially higher order. The necessity for this arises because  polynomial approximations of effective actions have a finite radius of convergence, often dictated by cuts or singularities in the complexified field plane \cite{Litim:2016hlb}. Therefore a reliable determination of de Sitter solutions necessitates a reliable determination of the domain of validity, primarily set by the radius of convergence. In addition, we will employ resummation techniques for the effective action including Pad{\'e} resummation and  numerical integration beyond polynomial orders. This allows us to investigate the existence (or not) of cosmological scaling solutions in the fixed point regime both for small and moderate Ricci curvature even beyond polynomial approximations.

The outline of our paper is as follows. After recalling the key features of inflationary scenarios and gravitational renormalisation group equations, we analyse  the fixed point solutions, the radius of convergence of high order polynomial approximations,  convergence-limiting singularities in the complex field plane, and compare findings with resummations and numerical integration 
(Sect.~\ref{QGandinfla}). We then exploit our findings to identify stationary  solutions of the effective action, reliably, both for small and large Ricci curvature. The domain of validity is critically assessed. We also perform resummations for the equation of state to discuss de Sitter and near de Sitter solutions as well as the impact of higher order invariants  (Sect.~\ref{stat}). We close with some conclusions
(Sect.~\ref{C}).

\section{\bf Asymptotically safe gravity and inflation}
\label{QGandinfla}
In this section, we discuss quantum gravity effects for inflation. Our main emphasis relates to the high-energy limit and the regime where gravity displays an asymptotically safe fixed point under the renormalisation group.

\subsection{Inflation and quantum gravity}\label{QG}
Inflation is the theory of the universe for which space-time undergoes a phase of accelerated expansion. There are strong observational indications for inflationary phases,  both during the early-time \cite{Ade:2013uln,Ade:2013zuv} and the late-time cosmological evolution of the  universe \cite{Perlmutter:1998np}.  
During an inflationary era of the universe the spacetime metric is approximately that of a de Sitter universe. 
Simple models of inflation include single scalar field theories coupled to gravity. Inflation may also be generated purely gravitationally. In either case,
accelerated expansion sets in provided the scalar curvature $R$ becomes nearly constant so that the dynamical evolution of the universe is very similar 
to that of a de Sitter universe. 
To be specific, we consider a FRW universe with scale parameter $a(t)$ and a gravitational action of the $f(R)$ type. The scalar curvature is then related to the Hubble parameter $H=\dot{a}/a$  by $R=12(H^2-\epsilon)$, where $\epsilon=-\dot{H}/H^2$
is the slow-roll parameter, and an overdot denotes a derivative wrt cosmological time. For sufficiently small $\epsilon$, $\dot\epsilon$ an exponential expansion is observed with $a\propto e^{H\,t}$. In this case, the equation of motion becomes 
\beq \label{EofM}
R f'(R) - 2 f(R) = 0
\eeq
whose solutions $R=R_{\rm dS}$ determine the Hubble parameter. 
An effective $f(R)$ theory could have two different types of de-Sitter solutions:
One which would end after a finite time with $R_{\rm dS}= R_{\rm infl}$ and $f(R) \propto R^2_{\rm infl}$ leading to inflation in the early universe \cite{Starobinsky:1980}, and a second one for which $R_{\rm dS} = 4 \Lambda$ (where $\Lambda$ is the cosmological constant) giving rise to 
accelerated expansion when dominating the contributions to the energy density of the universe.

From the perspective of quantum gravity, effective actions of the $f(R)$ type can arise due to quantum fluctuations of the metric field.
A possible window into the form of such actions comes from the asymptotic safety scenario for quantum gravity, which stipulates that the UV action should be a fixed point of the renormalisation group.   Here we shall assume that the form of such an action is determined by the existence of an asymptotically safe fixed point for a quantum version of gravity. 
The main new effect is that the gravitational couplings become running couplings depending on the renormalisation group scale $k$. 
The most widely studied actions which have been considered for asymptotic safety are those of the $f(R)$ form 
\cite{Codello:2007bd,Codello:2008vh,Machado:2007ea,Falls:2013bv,Falls:2014tra}. 
For lower energies and curvatures the action should approximate the Einstein-Hilbert form with a small but possibly non-zero cosmological constant which could provide for late time acceleration.

A suitable de Sitter fixed point leading to a viable inflationary model has been found in an $f(R)$-type approximations not including a cosmological constant \cite{Copeland:2013vva}.  Additional scalar matter fields have been taken into account in \cite{Contillo:2011ag,Hindmarsh:2011hx,Ahn:2011qt,Cai:2011kd,Hindmarsh:2012rc,Kaya:2013bga,Saltas:2015vsc}.
 In this paper, we will instead include a cosmological constant and we will search for de Sitter points based on the high order results in \cite{Falls:2013bv,Falls:2014tra}. 
In this way, our analysis also covers earlier approaches on cosmological models with asymptotic safety based on pure gravity as studied in \cite{Weinberg:2009wa,Reuter:2005kb,Bonanno:2001xi,
Bonanno:2001hi,Bonanno:2009nj,Bonanno:2010bt,Bonanno:2012jy,
Frolov:2011ys,Cai:2011kd,Hindmarsh:2012rc,Koch:2010nn}.

\subsection{Quantum gravity and the renormalisation group}\label{RG}
Next we recall how a function $f(R)$ is obtained from quantum gravity effects. We begin with gravitational effective actions in four-dimensional euclidean space-time  which are functions of the Ricci scalar
\beq\label{S}
\Gamma_k=\int d^4x \sqrt{|\det g_{\mu\nu}|}
\ F_k(\bar R)\,.
\eeq
where $\bar R=\bar R(g_{\mu\nu})$ denotes the Ricci scalar and $g_{\mu\nu}$ the metric field. The index $k$ denotes the renormalisation group (RG) momentum scale. Effective actions which are generic functions of the curvature scalar are of interest for cosmological model building and dark energy, see \cite{DeFelice:2010aj}. 
In our setup, however, the form of the function $F_k$ is not an input: Rather, its shape is entirely dictated by quantum gravity and must be determined by explicit computation.
In more concrete terms, the RG scale dependence of \eq{S} is governed by its functional RG flow  \cite{Wetterich:1992yh,Reuter:1996cp},
 \begin{equation}\label{dGamma}
 \partial_t\Gamma_k=\frac12 \Tr\frac{1}{\Gamma^{(2)}_k+R_k}\partial_t R_k\,.
 \end{equation}
Here, $R_k$ denotes a suitably defined Wilsonian momentum cutoff \cite{Litim:2000ci,Litim:2001up,Litim:2001fd}, $t = \ln k$, and $\Gamma^{(2)}_k$ the variation of the effective action with respect to the propagating fields. 
The gravitational effective action \eq{S} interpolates between an asymptotically safe fixed point action $\Gamma_*$ in the limit $k\to\infty$ and a semi-classical low-energy effective action $\Gamma_0$ in the limit $k\to 0$. For the latter, the effective action should fall back onto the Einstein-Hilbert action
\beq
F_0(\bar R)\approx \frac{\Lambda}{8\pi\,G}-\frac{1}{16\pi G} \bar R+\cdots\,,
\eeq
possibly up to higher order corrections in the Ricci scalar. Here, $\Lambda$ denotes the  cosmological constant, $G=1/(8 \pi\, M_{\rm Pl}^2)$ is Newton's constant  with  $M_{\rm Pl}=2.4.35 \times10^{18}$~GeV the reduced Planck mass, and ${\Lambda}/(8\pi\,G)\approx 10^{-47}~{\rm GeV}^4$ the vacuum energy. 
The perturbative iterative solution of \eq{dGamma} reproduces the conventional loop expansion \cite{Litim:2001ky,Litim:2002xm}. In this limit, corresponding to weak gravitational interactions $G \cdot k^2 \ll 1$, the perturbative non-renormalisability of gravity is recovered. 
A key feature of functional renormalisation with \eq{dGamma}  is that it does not necessitate weak coupling. Rather, it allows systematic approximations even at strong coupling where gravitational interactions become of order unity. 
The main physics novelty to be exploited below  is that the running Newton coupling  $g(k)=G(k) k^2$  does not diverge in the UV, but rather takes a finite value $g\to g_*$ owing to its own quantum fluctuations. This is the phenomenon of asymptotic safety.
The gravitational fluctuations also dictate the shape of the function $F_k$ in \eq{S}.

We add a few technical remarks and refer to
 \cite{Codello:2007bd,Codello:2008vh,Machado:2007ea,Benedetti:2012dx} for more details.
 The RG flow for $F_k$ is obtained by inserting \eq{S} into \eq{dGamma}, together with  suitable  gauge fixing and Faddeev-Popov ghost terms. To find explicit expressions, 
the operator trace on the RHS of \eq{dGamma} for the action \eq{S} is evaluated using background field techniques. The Hessian is obtained by expanding the second variation of the action around a suitable background metric
in the form of $g_{\mu\nu}=\bar{g}_{\mu\nu}+h_{\mu\nu}$. Gauge invariance is ensured using the background field method. The operator trace is evaluated using the early-time heat kernel expansion and spherical backgrounds with constant scalar curvature. Intermediate  technical steps simplify when using the York decomposition and  optimised momentum cutoffs \cite{Litim:2000ci,Litim:2001up,Litim:2003vp}.
Analysing the flow equation for  $F_k$ within a power series in the curvature scalar shows the existence of an interacting UV fixed point with three relevant couplings related to the cosmological constant, Newton's coupling, and $R^2$ interactions. The scaling dimensions for these deviate from Gaussian values owing to large anomalous dimensions. 
Furthermore, the fluctuations of the metric field are strong enough for gravity to become anti-screening towards shortest distances.
Higher order gravitational interactions also take interacting fixed points. For these, near-Gaussian scaling exponents are observed \cite{Falls:2013bv,Falls:2014tra}. 
As a final remark, it is useful to relate these findings with those obtained using conventional perturbation theory. Evaluating the flow equation \eq{dGamma}  perturbatively, and 
around  a flat background one recovers the results of Stelle \cite{Stelle:1976gc},  including the well-known issues with perturbative unitarity. In our setup, however, issues with massive ghosts are circumvented dynamically \cite{Benedetti:2009rx} owing to the spectral positivity of the perturbative RG flow, see \cite{Niedermaier:2009zz,Niedermaier:2010zz}.
Also, the notorious Goroff-Sagnotti term which plays a central role for gravity's perturbative non-renormalisability becomes irrelevant once gravity becomes anti-screening \cite{Gies:2016con}.

\subsection{Quantum gravity in the fixed point regime}\label{FPregime}

From now on we will investigate the shape of the function $F$ in the deep UV regime of the theory.
The explicit form of the fixed point action $\Gamma_*$, and hence the function $F$, has been derived within high-polynomial orders in the Ricci scalar  in \cite{Falls:2013bv,Falls:2014tra}. It is the central object of this study. 
For the remainder, it is convenient to introduce dimensionless fields, and we write
\begin{equation}\label{Rf}
\begin{array}{rcl}
R&=&\bar R/k^2\\[.5ex]
f(R)&=&16\pi\, F(\bar R)/k^4\,.
\end{array}
\end{equation}
Following  \cite{Falls:2013bv,Falls:2014tra}, we found it convenient to introduce an extra factor $16\pi$ into the definition of $f$, ensuring that Newton's coupling is given as $-1/f'$ in units of the RG scale without additional numerical factors, $G\cdot k^2=-1/f'$.
  Explicit functional flows \eq{dGamma} for actions \eq{S} in four euclidean dimensions have been given in \cite{Codello:2007bd,Codello:2008vh,Machado:2007ea}, and in \cite{Benedetti:2012dx} based on the on-shell action, also using \cite{Litim:2000ci,Litim:2001up,Litim:2003vp}. To facilitate consistency checks and a comparison with earlier findings we have adopted the approach put forward in \cite{Codello:2007bd,Codello:2008vh}.  
We are interested in the shape of $f(R)$ in the deep UV limit where the theory displays an interacting UV fixed point. At a non-trivial fixed point, the function $f(R)$ becomes scale-invariant, 
\beq\label{dtf}
\partial_t f(R)=0\,.
\eeq 
Using the explicit functional RG equations as discussed in the previous section, the fixed point condition \eq{dtf} leads to an explicit, non-linear  differential equation determining the function $f(\R)$ \cite{Codello:2007bd,Codello:2008vh} 
\begin{eqnarray}
\frac{df''(R)}{d\R}&=&
\frac{\frac{37}{756} \R^4+\frac{29}{10} \R^3+\frac{121}{5} \R^2+12 \R-216}{\frac{181}{1680}\R^4+\frac{29}{15} \R^3+\frac{91}{10} \R^2-54}\,\frac{f''(\R)}{\R}
-
\frac{\frac{37}{756} \R^3+\frac{29}{15} \R^2+18 \R+48}{\frac{181}{1680} \R^4+\frac{29}{15} \R^3+\frac{91}{10} \R^2-54}
   \,\frac{f'(\R)}{\R}
   \nonumber\\[1.5ex] &+&
\frac{(\R-3)^2 f''(\R)+(3-2 \R) f'(\R)+2 f(\R)}{\R \left(\frac{181}{1680} \R^4+\frac{29}{15} \R^3+\frac{91}{10} \R^2-54\right)}\times
\label{df''}
\\ &\times&
\left[\frac{\R \left(-\frac{311}{756} \R^3+\frac{1}{6}\R^2+30 \R-60\right)
   f''(\R)+\left(\frac{311}{756} \R^3
     -\frac{1}{3}\R^2-90 \R+240\right) f'(\R)}{3 f(\R)-(\R-3) f'(\R)}
   \right.
   \nonumber\\ &&
   \nonumber
\left.-\frac{607 \R^2-360 \R-2160}{15
   (\R-4)}-\frac{511 \R^2-360 \R-1080}{30(\R-3)}+48 \pi  \big(\R f'(\R)-2  f(\R)\big)\right]\,.
   \end{eqnarray}
An asymptotically safe  interacting fixed point solution $f_*(R)$ with \eq{dtf} and \eq{df''} has been given in \cite{Falls:2013bv,Falls:2014tra}. Its ``UV critical surface" is found to be three-dimensional. The relevant couplings are mainly given by  the vacuum energy, Newton's coupling, and the $R^2$ coupling, in the sense that adding further invariants to the polynomial approximation does not lead to new relevant directions. Moreover, all higher order polynomial couplings are irrelevant and approach near-Gaussian scaling exponents with increasing canonical mass dimension. Therefore, UV-safe trajectories running out of the fixed point are characterised by three fundamentally free parameters.

In the remaining part of the paper, we are interested in the UV fixed point solutions $f_*(R)$  to \eq{df''} and their main properties, both within polynomial expansions about vanishing curvature scalar, and within numerical solutions, based on integrations of \eq{df''} starting from suitable initial conditions.
We note that the RHS of \eq{df''} may become singular at specific points in field space arising through the various denominators in \eq{df''}, specifically 
\beq\label{singular}
\begin{array}{rcl}
\R&=&-9.99\,855\cdots\,,\\[.5ex]
\R&=&\ \ \,0\,,\\[.5ex]
\R&=&\ \ \,2.00\,648\cdots\,,\\[.5ex]
\R&=&\ \ \,3\,,\\[.5ex]
\R&=&\ \ \,4\,.
\end{array}
\eeq
The potential singularity at $R=0$ is harmless. Those at $R=3$ and $4$ are due to the momentum cutoff. In principle, fixed point solutions can be extended across such singularities at the expense of a free parameter, see \cite{Dietz:2012ic,Demmel:2015oqa,Ohta:2015fcu} 
for recent examples. This also signals the onset of a regime where the derivation of the RG flow based on the early time expansion of the heat kernel may no longer be trusted.  The potential singularities of the differential equation at $R=-9.99855$ and $R=2.00648$ arise from the scalar metric fluctuations. Therefore, for polynomial expansions   about $R=0$, we may expect from \eq{singular} that the radius of convergence $R_c$ is bounded from above by $R_{\rm max}$ given by
\beq\label{Rmax}
R_c\le R_{\rm max}=2.00648\cdots
\eeq
Extending fixed point solutions beyond \eq{Rmax} requires that the polynomial couplings at $R=R_c$ fullfill certain continuity conditions.

\begin{figure*}[t]
\begin{center}
\includegraphics[width=.7\hsize]{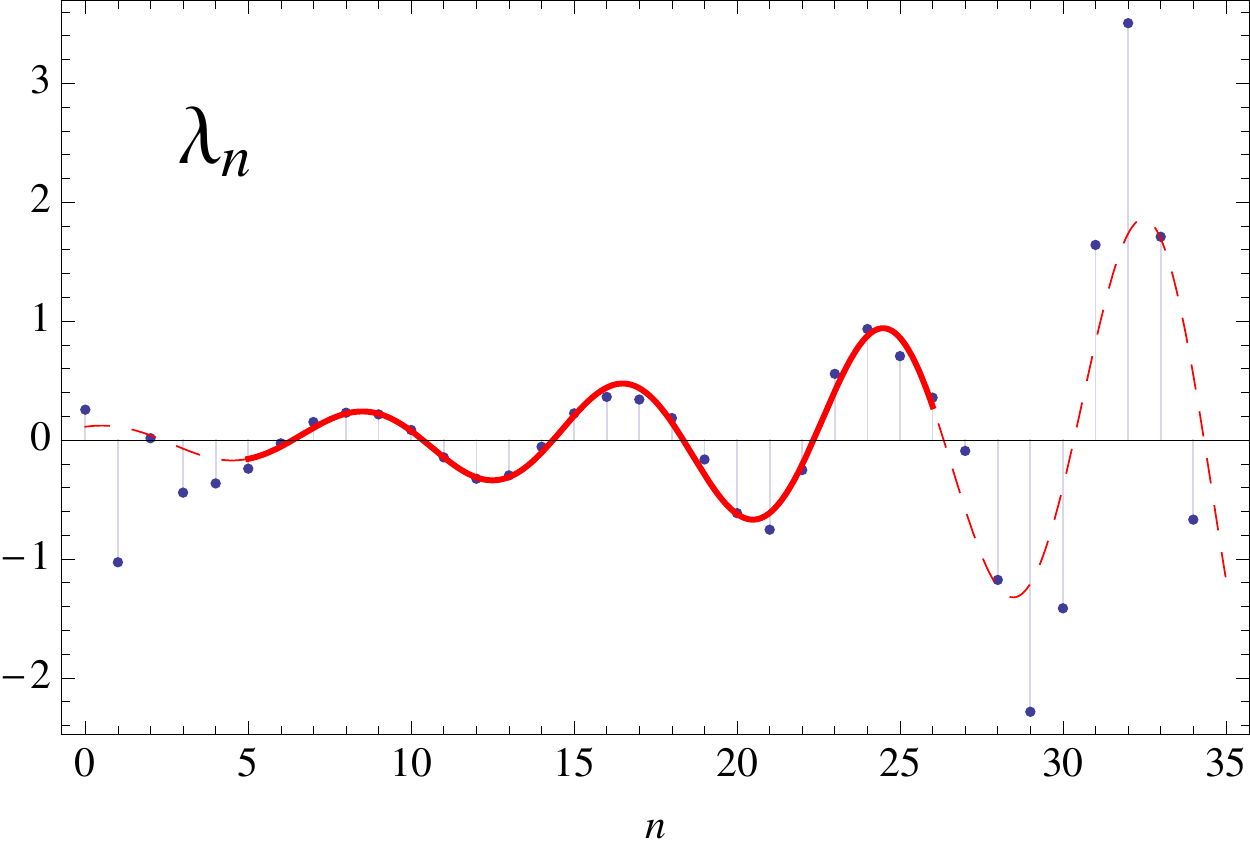}
\caption{\label{pCos} Comparison of the expansion coefficients $\lambda_n$ 
(dots) with the four-parameter fit \eq{cos}, \eq{fit1} (full red line) based on a selected set of couplings at approximation order $N=35$.
Dashed lines indicate extrapolations of the fit. 
}
\end{center}
\end{figure*}

\begin{center}
\begin{table}
\addtolength{\tabcolsep}{.1pt}
\setlength{\extrarowheight}{3pt}
\normalsize
\begin{tabular}{c | r r r r rrr r  } 
\hline\hline
\rowcolor{LightYellow}
$N$ & 35 & 31 & 27  & 23 & 19 & 15 & 11  & 7  \\\hline\hline
$\lambda_{0}$ & 0.25562 & 0.25555 & 0.25560 & 0.25546 & 0.25559 & 0.25522 & 0.25577 & 0.25388\\
$\lambda_{1}$ & $-$1.0272 & $-$1.0276 & $-$1.0276 & $-$1.0286 & $-$1.0281 & $-$1.0309 & $-$1.0289 & $-$1.0435\\
$\lambda_{2}$ & 0.01567 & 0.01549 & 0.01539 & 0.01498 & 0.01490 & 0.01369 & 0.01354& 0.007106\\
\rowcolor{LightGray}
$\lambda_{3}$ & $-$0.44158 & $-$0.44687 & $-$0.43997 & $-$0.44946 & $-$0.43455 & $-$0.45726 & $-$0.40246 &$-$0.51261\\
\rowcolor{LightGray}
$\lambda_{4}$ & $-$0.36453 & $-$0.36802 & $-$0.36684 & $-$0.37407 & $-$0.36981 & $-$0.38966 & $-$0.37114 &$-$0.48091\\
\rowcolor{LightGray}
$\lambda_{5}$ & $-$0.24057 & $-$0.23232 & $-$0.24584 & $-$0.23188 & $-$0.25927 & $-$0.22842 & $-$0.31678 &$-$0.18047\\
\rowcolor{LightGray}
$\lambda_{6}$ & $-$0.02717 & $-$0.02624 & $-$0.02286 & $-$0.01949 & $-$0.01564 & $-$0.002072 &$-$0.003987 & \cellcolor{LightRed} 0.12363\\
$\lambda_{7}$ & 0.15186 & 0.13858 & 0.15894 & 0.13620 & 0.17702 & 0.12649 & 0.23680 & $$\\
$\lambda_{8}$ & 0.23014 & 0.23441 & 0.22465 & 0.22904 & 0.21609 & 0.21350 & 0.23600 & $$\\
$\lambda_{9}$ & 0.21610 & 0.23820 & 0.20917 & 0.24918 & 0.18830 & 0.28460 & 0.12756 & $$\\
$\lambda_{10}$ & 0.08484 & 0.08207 & 0.092099 & 0.095052 & 0.095688 & 0.13722 & \cellcolor{LightRed} $-$0.041490 & $$\\
\rowcolor{LightGray}
$\lambda_{11}$ & $-$0.14551 & $-$0.17774 & $-$0.13348 & $-$0.19444 & $-$0.097057 & $-$0.25527 & \cellcolor{white}$$ &\cellcolor{white} $$\\
\rowcolor{LightGray}
$\lambda_{12}$ & $-$0.32505 & $-$0.33244 & $-$0.33242 & $-$0.36205 & $-$0.31812 & $-$0.46476 & \cellcolor{white}$$ &\cellcolor{white} $$\\
\rowcolor{LightGray}
$\lambda_{13}$ & $-$0.29699 & $-$0.25544 & $-$0.32410 & $-$0.24239 & $-$0.39520 & $-$0.16735 & \cellcolor{white}$$ & \cellcolor{white}$$\\
\rowcolor{LightGray}
$\lambda_{14}$ & $-$0.05608 & $-$0.04049 & $-$0.05633 & $-$0.000217 & $-$0.11204 & \cellcolor{LightRed} 0.16762 & \cellcolor{white}$$ & $$\cellcolor{white}\\
$\lambda_{15}$ & 0.22483 & 0.16347 & 0.26944 & 0.14317 & 0.37336 & $$ & $$ & $$\\
$\lambda_{16}$ & 0.36315 & 0.34000 & 0.37795 & 0.28611 & 0.50997 & $$ & $$ & $$\\
$\lambda_{17}$ & 0.34098 & 0.44488 & 0.28138 & 0.50187 & 0.17199 & $$ & $$ & $$\\
$\lambda_{18}$ & 0.18536 & 0.23941 & 0.15207 & 0.35074 & \cellcolor{LightRed} $-$0.11901 & \cellcolor{white}$$ & $$ &\cellcolor{white} $$\\
\rowcolor{LightGray}
$\lambda_{19}$ & $-$0.16304 & $-$0.32036 & $-$0.07588 & $-$0.41733 &\cellcolor{white} $$ &\cellcolor{white} $$ &\cellcolor{white} $$ &\cellcolor{white} $$\\
\rowcolor{LightGray}
$\lambda_{20}$ & $-$0.61457 & $-$0.73133 & $-$0.53776 & $-$0.95176 & \cellcolor{white}$$ &\cellcolor{white}  $$ &  \cellcolor{white}$$ &\cellcolor{white}  $$\\
\rowcolor{LightGray}
$\lambda_{21}$ & $-$0.75346 & $-$0.53875 & $-$0.88929 & $-$0.41230 & \cellcolor{white}$$ & \cellcolor{white} $$ & \cellcolor{white} $$ &\cellcolor{white}  $$\\
\rowcolor{LightGray}
$\lambda_{22}$ & $-$0.25160 & $-$0.05746 & $-$0.43756 & \cellcolor{LightRed} 0.29953 &\cellcolor{white} $$ &\cellcolor{white}  $$ &\cellcolor{white}  $$ & \cellcolor{white} $$\\
$\lambda_{23}$ & 0.55701 & 0.22998 & 0.73065 & \cellcolor{white}&\cellcolor{white}&\cellcolor{white}&\cellcolor{white}&\cellcolor{white}\\
$\lambda_{24}$ & 0.93392 & 0.60948 & 1.3116 & \cellcolor{white}&\cellcolor{white}&\cellcolor{white}&\cellcolor{white}&\cellcolor{white}\\
$\lambda_{25}$ & 0.70608 & 1.2552 & 0.54266 & &&&&\\
$\lambda_{26}$ & 0.35710 & 0.98891 & \cellcolor{LightRed} $-$0.31179 & &&&&\\
\rowcolor{LightGray}
$\lambda_{27}$ & $-$0.09106 & $-$0.92872 &\cellcolor{white}&\cellcolor{white}&\cellcolor{white}&\cellcolor{white}&\cellcolor{white}&\cellcolor{white}\\
\rowcolor{LightGray}
$\lambda_{28}$ & $-$1.1758 & $-$2.3752 & \cellcolor{white} & \cellcolor{white}  &\cellcolor{white}  & \cellcolor{white} & \cellcolor{white}  &\cellcolor{white}  \\
\rowcolor{LightGray}
$\lambda_{29}$ & $-$2.2845 & $-$1.1315 &\cellcolor{white}&\cellcolor{white}&\cellcolor{white}&\cellcolor{white}&\cellcolor{white}&\cellcolor{white}\\
\rowcolor{LightGray}
$\lambda_{30}$ & $-$1.4145 &\cellcolor{LightRed}  0.64746 &\cellcolor{white} &\cellcolor{white}&\cellcolor{white}&\cellcolor{white}&\cellcolor{white}&\cellcolor{white}\\
$\lambda_{31}$ & 1.6410 & &&&&&&\\
$\lambda_{32}$ & 3.5054 & &&&&&&\\
$\lambda_{33}$ & 1.7098 & &&&&&&\\
$\lambda_{34}$ & \cellcolor{LightRed} $-$0.66883 & &&&&&&\\
\end{tabular}  
\caption{\label{tFP} Coordinates of the ultraviolet fixed point $\lambda_n(N)$ in a polynomial base \eq{expansion0} for selected orders in the expansion. Note that $g_*=-1/\lambda_1$. We observe the approximate eight-fold periodicity pattern \eq{8} in the signs of couplings.  
Notice that the  couplings $\lambda_{2+4i}$ at approximation order $N_1=3+4i$ (the order at which they first arise) always come out with the ``wrong'' sign. To achieve the desired accuracy in  the subsequent analysis, we have retained at least 50 significant digits for the couplings. The data for  $N=7$  and $N=11$ agrees with  earlier findings in \cite{Codello:2007bd} and \cite{Bonanno:2010bt}, respectively. }
\end{table} 
\end{center}

\subsection{Polynomial fixed point and convergence}\label{FP}

Next we analyse the ultraviolet fixed point solution of polynomial $f(R)$ gravity of \cite{Falls:2013bv,Falls:2014tra} in four dimensions. Following \cite{Litim:2016hlb} we identify the location of a convergence-limiting pole (or cut) in the plane of complexified Ricci scalar, which is exploited to estimate the radius of convergence. In \cite{Falls:2013bv,Falls:2014tra}, we have computed the fixed point coordinates in a polynomial approximation up to order $N=N_{\rm max}=35$ in the expansion
\begin{equation}\label{expansion0} 
f(\R)=\sum_{n=0}^{N-1}\lambda_n \R^n\,.
\end{equation}
Numerical results for the fixed point couplings up to the order $N\le N_{\rm max}=35$ are summarized in Tab.~\ref{tFP}, showing the fixed point coordinates 
 for selected orders of the approximation. Notice that the signs of the couplings follow, approximately, an eight-fold periodicity in the pattern \cite{Falls:2013bv}
 \begin{equation}
 \label{8}
 (++++----)\,.
 \end{equation}
Four consecutive fixed point couplings $\lambda_{3+4i}-\lambda_{6+4i}$ come out negative (positive) for even (odd) integer $i\ge 0$. Periodicity patterns  such as this one often arise due to convergence-limiting singularities (poles or cuts) of the fixed point solution $f(\R)$ in the complex $\R$-plane, away from the real axis. If the  singularity nearest to the origin in field space arises at $R=R_0$ (which can and will be complex), then the radius of convergence for a field expansion about vanishing field is given by $R_{\rm c}=|R_0|$. This is well-known from scalar theories at criticality where  $2n$-fold periodicities are encountered regularly \cite{Morris:1994ki,Aoki:1998um,Litim:2002cf,Litim:2003kf,Litim:2016hlb}.  Below we determine $R_{\rm c}$ to establish properties of the interacting fixed point including the existence or not of de Sitter solutions.

\begin{figure*}[t]
\begin{center}
\unitlength0.001\hsize
\begin{picture}(1000,650)
\put(120,0){\includegraphics[width=.65\hsize]{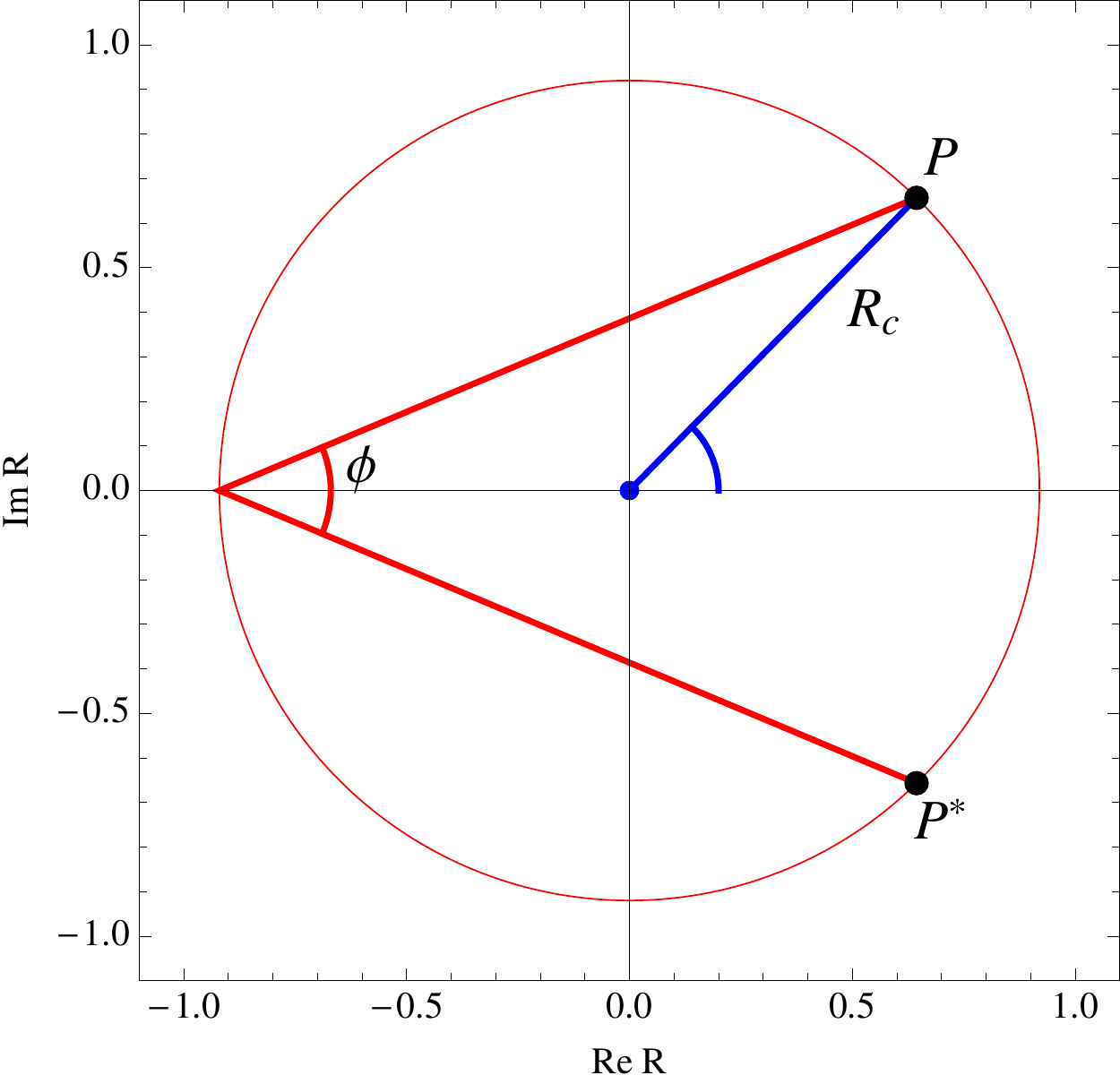}}
\end{picture}
\caption{\label{Poles} Location of the  convergence-limiting singularities (poles or cuts) of the ultraviolet fixed point solution $f_*(R)$  in the plane of complexified scalar curvature,
indicated by the black dots $P$ and $P^*$. Their approximate distance $R_c$ from the origin and the angle $\phi$ under which they appear are also indicated. Notice that $\phi$ is close to $\frac{\pi}{4}$ owing to  the eight-fold periodicity pattern \eq{8}.} 
\end{center}
\end{figure*} 

\label{sing}
The polynomial expansion \eq{expansion0} has a finite radius of convergence $R_c$, which is expected to arise due to a singularity or a cut of the fixed point solution for $f(R)$ in the complex field plane. Its location can be estimated from the convergence behaviour of the fixed point coordinates. Standard convergence tests such as the root test or the ratio test fail due to the eight-fold periodicity in the couplings, and a high-accuracy computation of $R_c$ requires many orders in the expansion. Therefore we adopt the following strategies. Firstly, we make a low-parameter fit of the fixed point couplings from Tab.~\ref{tFP} onto a form which reflects the observed periodicity pattern. This provides us with an estimate for $R_c$. Secondly, we then cross-check the findings with more sophisticated convergence tests such as a modified ratio test and the Mercer-Roberts test \cite{Litim:2016hlb}. 
We begin with a very simple four-parameter ansatz
\begin{equation}\label{cos}
\lambda_n= A\,\frac{\cos(n\,\phi+ \Delta)}{(R_c)^n} 
\end{equation}
for the couplings, motivated by the sign pattern observed in the couplings. Also, expansions such as \eq{cos} arise naturally provided that the fixed point displays a singularity in the complex plane.
We observe that the periodic pattern starts roughly from $n=5$ onwards, and hence we must omit the first few couplings for the fit. We also omit a few of the highest couplings $\lambda_n$ with $n>26$. Typically, the last cycle of eight highest couplings has not yet properly settled on their asymptotic values. For these reasons we fit \eq{cos} using the most reliable values of  the best solution $(N=35)$ given by approximately 2.5 cycles of data points (22 consecutive values from $\lambda_5$ up to $\lambda_{26}$). For the amplitude, the radius, the angle and the shift, we find
\begin{equation}\label{fit1}
\begin{array}{rl}
A&=\ \ \,0.1172\\[.5ex]
R_c&=\ \ \,0.9182\\[.5ex]
\phi&=\ \ \, 0.7863\\[.5ex]
\Delta&=-0.2919\,.
\end{array}
\end{equation}
Notice that the angle $\phi$
is very close to $\pi/4\approx 0.7854$, as expected due to the eight-fold periodicity pattern. 
Comparing with the exact fixed point values, we observe good agreement even beyond the fitted domain, see Fig.~\ref{pCos}.

A first cross-check is done using the Mercer-Roberts test based on the same data set. It  estimates the angle as $\phi=0.7736\pm 3\%$, confirming that \eq{fit1} provides a good estimate for the angle $\phi$ under which the fixed point solution displays a singularity in the complex field plane, see Fig.~\ref{Poles}.  The radius of convergence should be reliable on the 5-10\% level. Neither the root test, the ratio test nor  the Mercer-Roberts test provide stable results for the radius of convergence. As a cross-check for the radius, we therefore adopt   
a generalised ratio test according to which the radius is estimated via the limit
\beq\label{root}
R_c=\lim_{n\to\infty}\,
\left|\frac{\lambda_n}{\lambda_{n+m}}\right|^{1/m}
\eeq
provided it exists, irrespective of the free parameter $m$.
It turns out  that if $m$ is taken to be  the underlying periodicity or larger, $m\ge 8$, the ratios $\left|{\lambda_n}/{\lambda_{n+m}}\right|^{1/m}$ depend only weakly on the parameter $m$, and converge well with increasing $n$. Since our data sets are finite, the limit $1/n\to 0$ can only be performed approximately. 
We obtain a mean value for $R$ by computing the ratios for various $m$, and then averaging over all $m$. In this manner, the estimate is as stable as it gets, and largely insensitive to the choice for $m$.  We find 
\begin{equation}\label{root}
R_c\approx 0.91\pm 5\%\,,
\end{equation}
and the statistical error of approximately a standard deviation is due to the variation with $m$.
\begin{figure*}[t]
\begin{center}
\unitlength0.001\hsize
\begin{picture}(1000,500)
\put(110,-10){\includegraphics[width=.75\hsize]{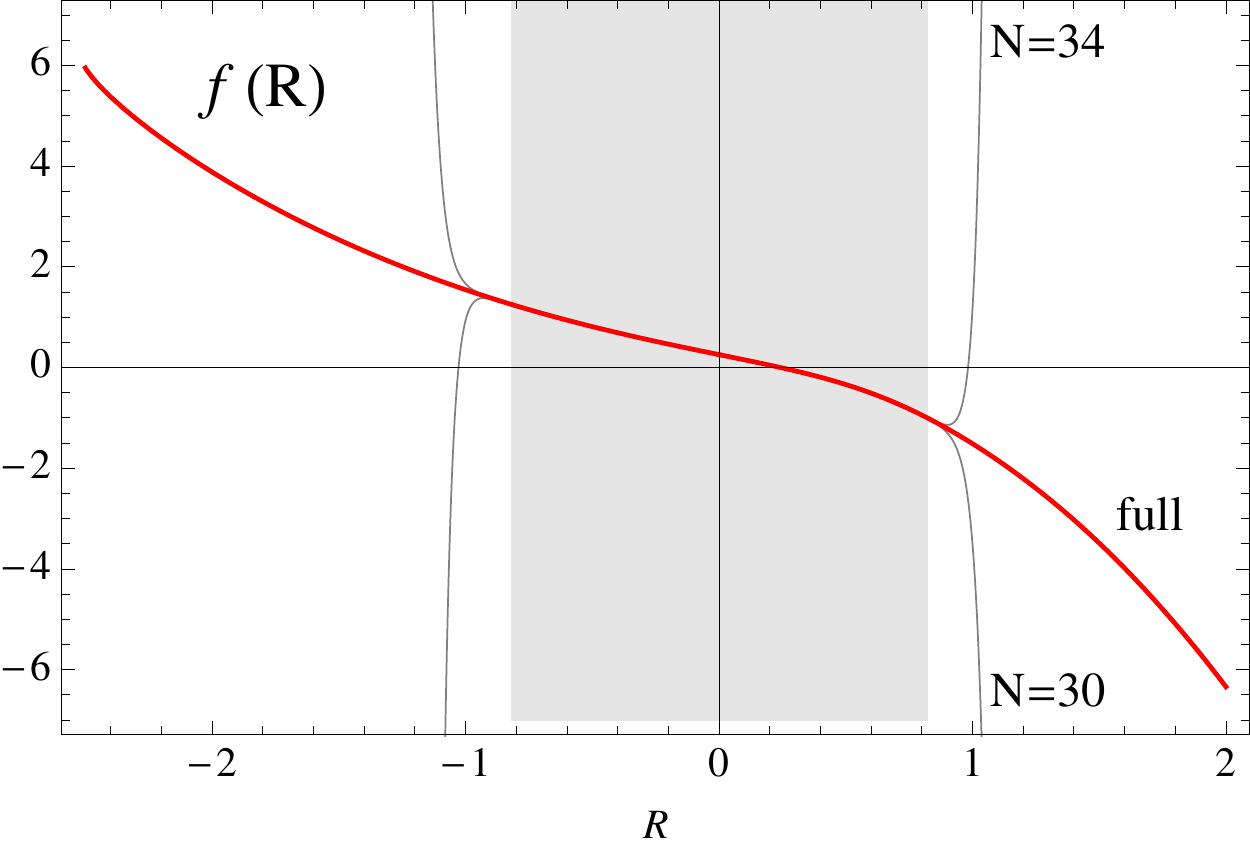}}
\end{picture}
\caption{\label{ExpansionAtHighestOrder}The  field-dependent UV fixed point $f(\R)$, comparing the full numerical integration of \eq{df''} (no polynomial approximation; full red line) with the polynomial approximation $f_N$ up to orders $N=30$ and $34$ (thin gray lines). The shaded area extends up to the radius of convergence \eq{RL}. }
\end{center}
\end{figure*}Notice that this procedure is insensitive to the angle $\phi$. It is therefore interesting that the estimates \eq{root} and  \eq{fit1} agree.  The smallness of the statistical error indicates that the radius is achieved for most of the data points except for a few where the radius comes out smaller. In this light, a conservative `lower bound' for the radius is then obtained by projecting onto the smallest estimate 
\beq\RLB(m)\equiv\min_n \left|\frac{\lambda_n}{\lambda_{n+m}}\right|^{1/m}
\eeq 
from the most advanced data set $(N=35)$, and  for each admissible parameter $m$ $(8\le m\le N-m)$. We then average these lower bounds over $m$
to estimate the conservative lower bound as
\beq\label{RL}
\RLB\, \approx\,
0.82\, \pm\, 5\%
\eeq 
where, again, the statistical error of approximately a standard deviation is due to the variation with $m$. The smallness of the statistical error reflects that the value \eq{RL} is achieved for essentially all $m\ge 8$. 
For illustration, we show in Fig.~\ref{ExpansionAtHighestOrder}  the fixed point solution as a function of $R$ to order $N=31$ and $N=35$. Both solutions visibly part  each other's ways at fields of the order of \eq{RL}, supporting our rationale.

We now relate our findings with earlier studies of the radius of convergence  based on approximations up to order $N=11$ \cite{Bonanno:2010bt}. There, a larger radius of convergence has been found to be
\begin{equation}\label{N11}
R_c\approx 0.99
\end{equation} 
adopting a procedure different from ours. Error estimates have not been given. Had we restricted our procedure to the first 11 fixed point couplings  (by using either the $N=11$ data, or the first 11 entries from the $N=35$ data set), our analysis leads to 
\begin{equation}\label{N11new}
R_c\approx 1.0\pm 20\%\,. 
\end{equation}
Within errors, \eq{N11new} is consistent with the more accurate estimates \eq{root} and \eq{RL}. It is also consistent with the earlier estimate \eq{N11}. 
The  slight over-estimation for $R_c$ at low orders \eq{N11new} can be understood in terms of the eight-fold periodicity pattern underlying the data. A first full cycle of eight is completed  at approximation order $N=11$; see Tab.~\ref{tFP}. At approximation order $N=35$, four such cycles have been completed, offering a more accurate estimate of the radius of convergence.

We also note that the various estimates for $R_c$ \eq{fit1},\eq{root} and the lower bound $R_{L}$ \eq{RL} are substantially smaller than the formal upper bound $R_{\rm max}$ identified in \eq{Rmax}. Hence, the theoretically achievable radius of convergence is not realised by the fixed point solution. Rather, a singularity or cut in the plane of complex Ricci scalar imposes more stringent limits.

\subsection{Fixed point beyond polynomial orders}
Given the good convergence of fixed point couplings, our results should not vary strongly if the polynomial expansion is pushed to higher orders, including the asymptotic limit $N\to \infty$. Furthermore, as we argued, the polynomial expansion has a finite radius of convergence, and pushing the expansion towards $N\to \infty$ will not provide  access to the regime $|R|>R_c$. Therefore, in order to go beyond all polynomial orders, we must find the fixed point solution beyond $\R_c$ by integrating the fixed point condition numerically with initial data provided by the polynomial approximation, see Fig.~\ref{ExpansionAtHighestOrder}. Since the fixed point condition is a third order differential equation for $f$, \eq{df''}, we need to give three initial conditions. At $\R =0$ this reduces to two initial conditions since one condition is ``used up'' in order 
 to avoid a divergence at the origin. This leaves us with the two free parameters $\lambda_0$ and $\lambda_1$. To numerically integrate \eq{df''} 
into positive (negative) values of $\R$ we take initial conditions for $f$, $f'$ and $f''$ from our highest polynomial approximation $(N=35)$ 
at field values $\R = 0.1$ ($\R = -0.1$) deeply within the radius of convergence. We have checked that our results for $f$ are independent of these technical choices.

In Fig.~\ref{ExpansionAtHighestOrder} we compare the full numerical integration  (thick red curve) with  the polynomial approximations at order $N=34$ and $N=30$ (thin gray curves) and note that we are able to compute $f$ outside the radius of convergence $\R_L$. 
The polynomial and numerical solution coincide within the radius of convergence, as they must. The validity of the full numerical solution extends substantially beyond $R_L$. However, we cannot integrate the fixed point up to infinite field due to technical singularities  of the differential equation \eq{df''} at intermediate or large curvature scalar. With increasing $\R$, the first integrable pole is located at $\R\approx 2.006$, see \eq{singular}. On the negative curvature axis, the closest integrable pole is located at $\R\approx-9.99$. Both of these can in principle be dealt with using ideas discussed in eg.~\cite{Dietz:2012ic}.  Our numerical integration becomes unreliable close to $R\approx 2.006$ and close to $R\approx -2.541$,  which prevents the solution being continued to larger values of $|R|$.  For asymptotically large $R$, the approximations adopted in \eq{df''} are less reliable, and we can expect corrections in a more complete treatment.

\subsection{Resummation}\label{resum}

       \begin{figure*}[t]
\begin{center}
\unitlength0.001\hsize
\begin{picture}(1000,500)
\put(110,-10){\includegraphics[width=.75\hsize]{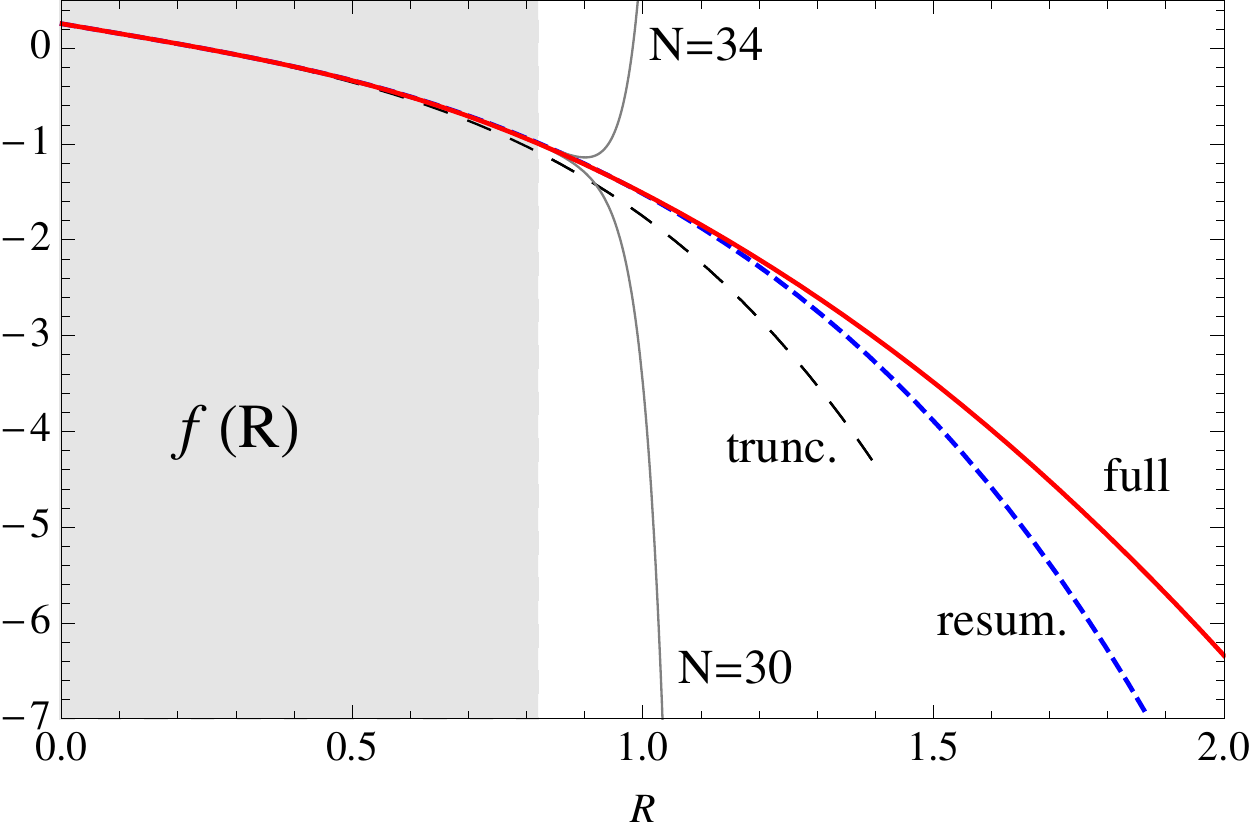}}
\end{picture}
\caption{\label{pResum}Shown are results for the  field-dependent fixed point $f(\R)$, comparing the polynomial approximations $f_N$ to orders $N=30$ and $34$ (thin gray lines) with the resummation $f_{\rm resum.}$ (dashed blue line) and the truncation $f_{\rm trunc}$ (long-dashed line) detailed in \eq{resum}, \eq{resum2}, and the full numerical integration (full red line). The shaded area indicates the region up to the radius of convergence of the polynomial approximation. We observe that the  resummation offers a substantial improvement over $f_{\rm trunc.}, f_{30}$ and $f_{34}$. 
It agrees with the high-order polynomials within,
and with the numerical integration beyond, the radius of convergence.}
\end{center}
\end{figure*}

The results of the previous sections show that the polynomial fixed point solution is bounded to the regime of small fields. However, one may wonder whether  resummation techniques could be used to extrapolate results reliably towards larger values for the scalar curvature. 
We begin by exploring whether the eightfold periodicity in the result \eq{fit1} can be used to resum the polynomial series \eq{expansion0} to infinite order. To that end we write
\beq\label{resum}
f_{\rm resum.}(R)=
\sum_{n=0}^{m-1}\lambda_n\,R^n+\sum_{n=m}^\infty\lambda_n\,R^n\,.
\eeq
Fig.~\ref{pCos} indicates that the eightfold periodicity pattern sets in at about $n\approx 5$. Therefore we take $m=5$ in \eq{resum}. We then leave the first sum as it is, and denote it as $f_{\rm trunc}(R)$. The second sum can be evaluated in closed form if we assume that the couplings $\lambda_n$ are well-approximated by \eq{cos} for all $n>4$. This leads to the resummed expression
\beq\label{resum2}
\sum_{n=5}^\infty\lambda_n\,R^n=A\,\frac{R^4}{R_c^4}\,\frac{R\,R_c\cos(\Delta+5\phi)-R^2\,\cos(\Delta+4\phi)}{R^2-2\,R\,R_c\cos\phi+R_c^2}
\eeq
in terms of four parameters $A, R_c,\phi$ and $\Delta$ which are, approximately, given by \eq{fit1}.
Owing to the split \eq{resum}, the resummed expression behaves $\propto R^4$ asymptotically. 

In Fig.~\ref{pResum}, we compare  the resummation \eq{resum}, \eq{resum2} (short dashed blue line) with the polynomial approximation at high orders ($N=30,34$: thin gray lines), the truncation $f_{\rm trunc.}$ ($N=5$: long dashed black line), and the full numerical integration (full red line). The difference between the long-dashed and the short dashed line corresponds to the resummed terms \eq{resum2}. While $f_{\rm trunc}$ starts deviating from the exact result already for $R$ below the radius of convergence $R_L\approx 0.82$, we observe that $f_{\rm resum.}$ provides a much better approximation for $f(R)$ even beyond $R_L$ of the polynomial expansion. The difference $f_{\rm resum}(R)-f_{\rm trunc}(R)$ is only sensitive to the location of the singularity nearest to the origin: the information initially encoded in all the higher couplings $\lambda_n$ with $n>4$ has been reduced to four parameters (concretely, here, 26 parameters have been reduced to four). This input appears to be enough to bring $f_{\rm resum}$ quite close to the solution from the full numerical integration, up to values of the curvature $|R|\approx 1.1$, and clearly beyond the domain of validity $R_L\approx 0.82$ of the polynomial approximation. We conclude that the simple four-parameter resummation \eq{resum2}  offers a substantial improvement.

\begin{figure*}[t]
\begin{center}
\unitlength0.001\hsize
\begin{picture}(1000,500)
\put(100,-10){\includegraphics[width=.75\hsize]{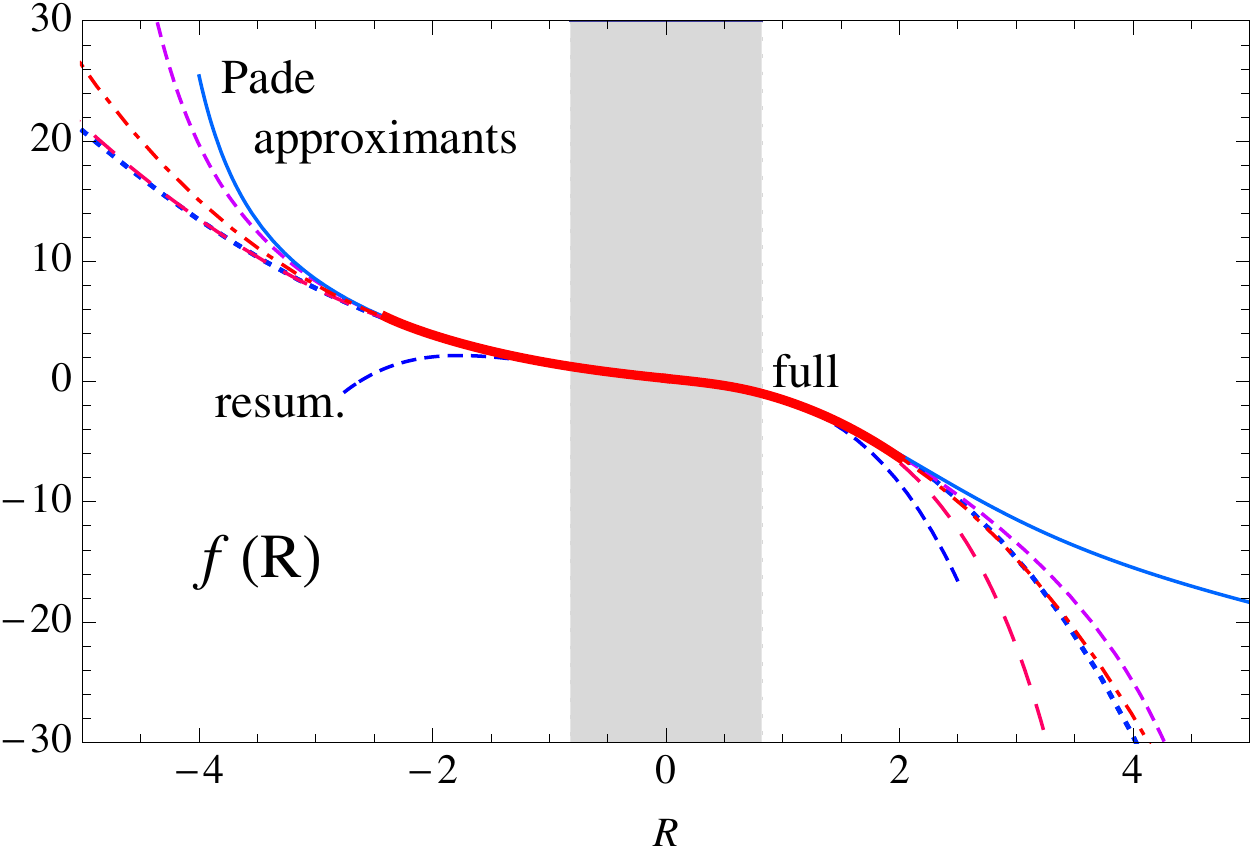}}
\end{picture}
\caption{\label{pResumPade} 
Shown is the field-dependent function $f(R)$ at the fixed point obtained by full numerical integration of \eq{df''}  (thick red line) in comparison with  Pad\'e approximants thereof  ([20/16]: dotted line, [17/14]: dashed-dotted, [15/13]: long-dashed, [16/15]: thin  full, [16/17]: short dashed), and the resummation \eq{resum}, \eq{resum2} (resum./dashed line). The shaded area indicates the radius of convergence $R_L\approx 0.82$ of the underlying polynomial approximation  \eq{expansion0}. Despite of differences in their asymptotic behaviour, the Pad\'e approximants agree very well amongst themselves and with the full numerical solution  for scalar curvatures within $-2.5\lta R\lta 2$ (they coincide with the full red line in  Fig.~\ref{ExpansionAtHighestOrder}), and way beyond the radius of convergence.  See main text for more detail.
}
\end{center}
\end{figure*}

\subsection{Pad\'e approximants}\label{pade}

Motivated by the findings of the previous section, we now adopt the more sophisticated technique of Pad\'e approximants. A Pad\'e approximant of the function $f(R)$ in \eq{expansion0} is a rational function -- the ratio of two unique polynomials in $R$ of degree $M$ and $K$ -- which we denote as $[M/K]$. The integers $M\ge 0$ and $K\ge 1$ are at our disposal. The first $M+K+1$ Taylor coefficients of $f(R)$ serve as input, and, by construction, agree with those of the approximant $[M/K]$.  Unlike in \eq{resum2}, the number of parameters is not reduced, and, consequently, we expect that a successful Pad\'e resummation leads to a further improvement upon $f_{\rm resum.}(R)$.

In practice, and in order to exploit the maximum number  of Taylor coefficients $N_{\rm max}=35$, we take $M$ close to and below $N_{\rm max}/2$. This is combined with  $K=M+1,M-1,M-2,M-3,M-4$ to account for different large-$R$ asymptotics.  Concretely, in Fig.~\ref{pResumPade} we show a selection of best fits corresponding to Pad\'e approximants $[{M}/{K}]$ with  $M=16$, $K=M+1$ (short-dashed line); $M=16$, $K=M-1$ (full line); $M=15$, $K=M-2$ (long-dashed line); $M=19, K=M-3$ (dashed-dotted line); and $M=20, K=M-4$ (dotted line). 

Three comments are in order. Firstly, it is quite noteworthy that all Pad\'e approximants not only match each other within the radius of convergence of the polynomial expansion \eq{RL},
but also  in the much wider regime $-2.5\lta R\lta 2$, substantially beyond the radius of convergence of the polynomial approximation. 
Secondly, the Pad\'e approximants also coincide with the full numerical solution of Fig.~\ref{ExpansionAtHighestOrder} within the same range $-2.5\lta R\lta 2$. 
The domain of validity of the resummed result is more than twice  as large as the original one estimated in \eq{RL}. 
Finally, for even larger fields the different resummations deviate from each other owing to the differences in the assumed asymptotic behaviour. These set in earlier, and are more pronounced, for positive than for negative curvature. Presumably this is so because the nearest singularity on the real axis of the fixed point differential equation \eq{df''} arises first for positive curvature, see \eq{singular}. 

We conclude that suitably adapted resummation techniques are powerful tools to increase the domain of validity for polynomial fixed point solutions. The polynomial couplings $\lambda_n$ appear to encode information about $f(R)$ beyond $R_L$, possibly up to the maximal range \eq{Rmax}.  The fixed point data is not sufficient to infer the large-field asymptotic behaviour of the function $f(R)$.  This should not come as a surprise:  heat kernel expansions have been used in the first place to obtain the fixed point coordinates which is a very good approximation for small, but less so for large scalar curvature. 

\section{\bf De Sitter and near-de Sitter points}\label{stat}

In this section, we apply the findings of the previous section to analyse stationary solutions of fixed point actions, which are of interest for the cosmological evolution in the early universe. Subject to suitable matching conditions relating the RG scale parameter with physical scales such as the Hubble parameter, or others,  the modified cosmological equation may lead to modified Friedmann equations with inflationary regimes \cite{Shapiro:2000dz,Bonanno:2001xi,Weinberg:2009wa,Bonanno:2010bt,Contillo:2010ju,
Hindmarsh:2011hx,Cai:2011kd,Contillo:2011ag,Hindmarsh:2012rc,Copeland:2013vva}. The availability of stationary points within asymptotically safe $f(R)$ gravity has been investigated previously in  \cite{Bonanno:2010bt} up to order $N=11$ in a polynomial approximation. More recently stationary points have been found in studies of asymptotically safe $f(R)$ gravity utilising an exponential parameterisation of the metric \cite{Ohta:2015efa,Ohta:2015fcu}, or a modified functional measure \cite{Demmel:2015oqa}. The importance of stationary points has been stressed in \cite{Dietz:2013sba} where it is suggested that the $f(R)$ approximation may break down in their absence.  Aspects of inflationary solutions in  higher derivative gravity are addressed in \cite{Maroto:1997aw}.

\begin{figure*}[t]
\begin{center}
\unitlength0.001\hsize
\begin{picture}(1000,500)
\put(91,-10){\includegraphics[width=.75\hsize]{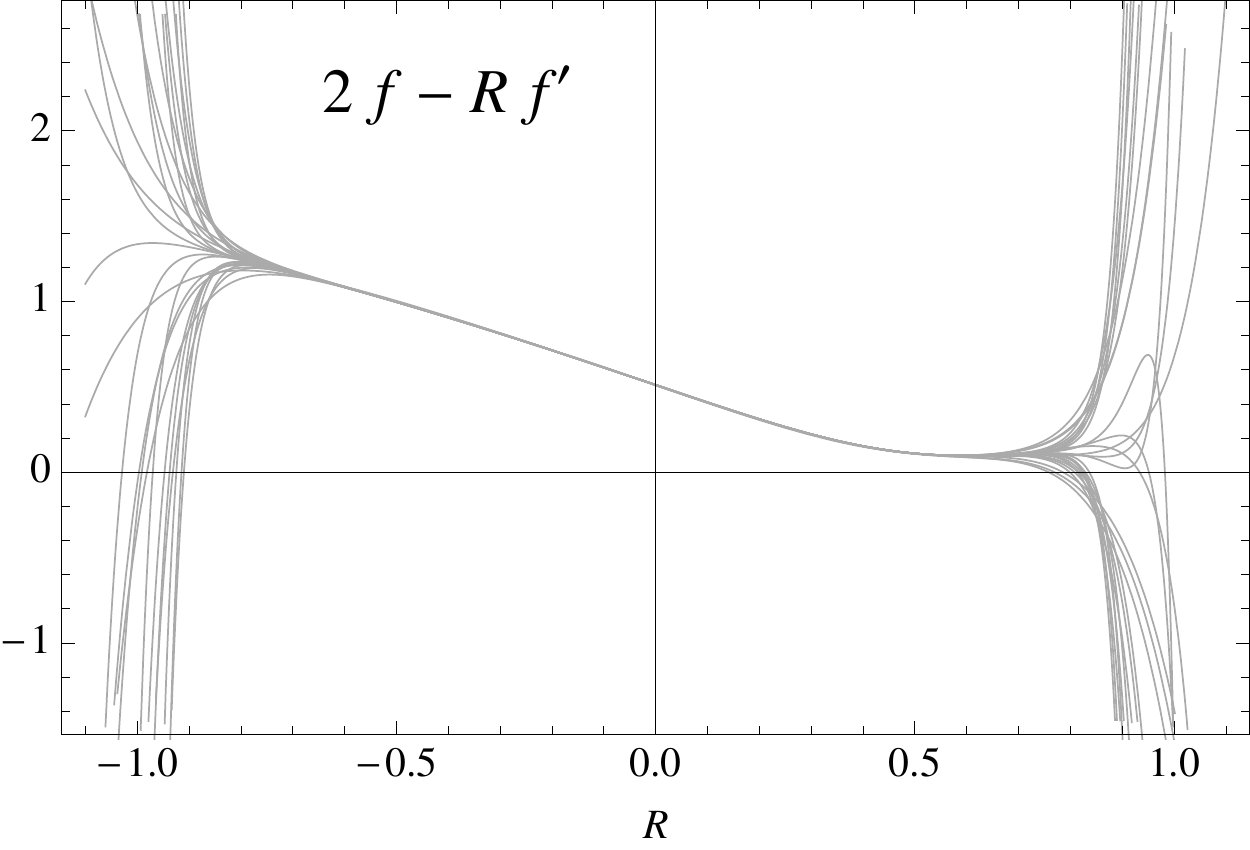}}
\end{picture}
\caption{\label{deSitterApprox}Shown is the stationarity condition for various orders in the polynomial approximation $9\le N\le 35$. The zeros, given in Tab.~\ref{tdS}, are candidate for real de Sitter or anti-de Sitter solutions.}
\end{center}
\end{figure*}

\subsection{Stationarity condition}

Given the RG fixed point to high order in the polynomial expansion, we now turn to the effective action which we write in terms of $f(R)$ as 
 \begin{equation} \label{Gammaf}
 \Gamma_k= \frac{k^4}{16 \pi}\int d^4x\sqrt{|\det g|}\,f_k(R)
\,. 
\end{equation}
Notice that the explicit RG scale dependence arises because we have used dimensionless fields and the dimensionless function $f$. Although we have computed $\Gamma_k$ on four-spheres, stationary points of constant curvature can be found on different topologies by analytical continuation of the metric and use of the appropriate boundary conditions. For Lorentzian metrics stationary points of constant (negative) positive curvature correspond to (anti-)de Sitter solutions to the quantum equations of motion for $g_{\mu \nu}$. Such solutions are given by the condition
\begin{equation}\label{deSitter}
\begin{array}{rl}
E(\R)&\equiv 2 f_k(\R)  -\R f_k'(\R)=0
\end{array}
\end{equation}
where we have introduced $E(\R)$ to denote the equation of motion related to the action \eq{Gammaf}. When evaluated on four-spheres however, the volume integration can be performed yielding
 \begin{equation}
 \int d^4x\sqrt{|\det g|}= \frac{384\pi^2}{R^2}\frac{1}{k^4}\,.
\end{equation}
Accordingly, we find
 \begin{equation}\label{Gammak}
 \Gamma_k=24\pi\frac{f_k(R)}{R^2}\equiv384\pi^2\frac{F_k(\bar R)}{\bar R^2}\,,
\end{equation}
where  we have re-introduced the action in terms of the dimensionful fields in the second equation. In this form, and for fixed $\R$ or $\bar R$, the sole RG scale dependence originates from the implicit $k$-dependence of the function $f$ or $F$, respectively.

\begin{figure*}[t]
\begin{center}
\unitlength0.001\hsize
\begin{picture}(1000,500)
\put(91,-10){\includegraphics[width=.75\hsize]{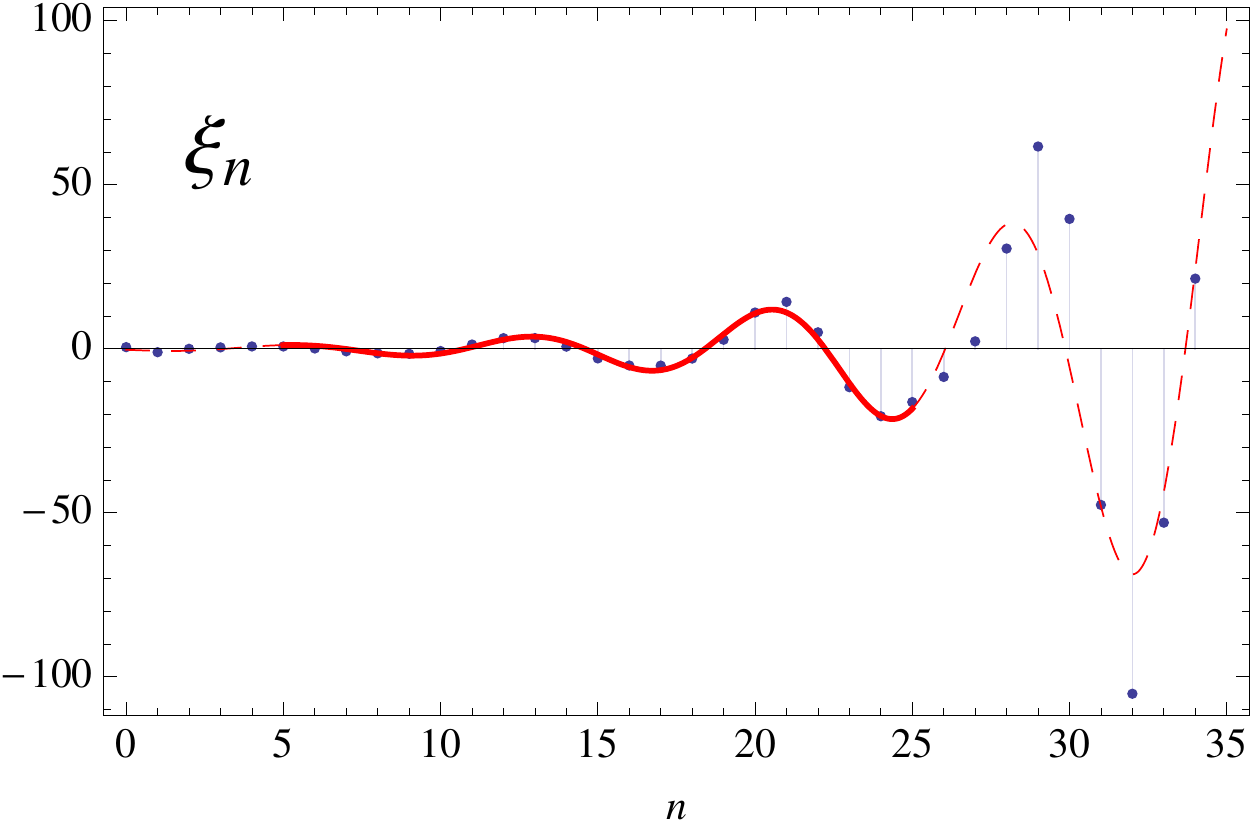}}
\end{picture}
\caption{\label{pdeSitterCos} Comparison of the expansion coefficients $\xi_n$ 
in \eq{zeta}  (dots) with the fit in \eq{desittercos} (full red line) based on a selected set of couplings at approximation order $N=35$.
The dashed line extrapolates the fit. 
}
\end{center}
\end{figure*}

Strictly speaking the function of the curvature given by \eq{Gammak}  is valid only on four-spheres. We also note that $\Gamma_*$ diverges at $R=0$, because the fixed point solution obeys $f(R=0)> 0$ whereas the volume diverges as $1/R^2$. To analyse the extremal points, we therefore must seperately discuss the cases of positive and negative curvature, and Euclidean and Lorentzian signatures. However, for curvatures $R\neq 0$ and $1/R\neq 0$, the stationarity for \eq{Gammak} 
\begin{equation}\label{condition1}
\frac{\partial\Gamma_k}{\partial \R}=0
\end{equation} 
 translates directly to the condition \eq{deSitter}. Therefore, assuming the validity of analytical continuation, we can use \eq{condition1} to find solutions to the quantum equations of motion.
Positive real solutions to \eq{deSitter} are termed de Sitter solutions $\R = \Rds$ in the literature. In a slight abuse of language, we will refer to any solution of \eq{deSitter} as $\R = \Rds$. The stationarity condition \eq{deSitter} remains algebraically the same at or away from fixed points
\footnote{Away from the fixed point, further contributions, for example by matter perturbations, will eventually become important enough to drive away from the de Sitter solution, see e.g. \cite{Mukhanov:1990me}.}, and it holds equally for dimensionful fields and the dimensionful function $F$ after trivial replacements.  It is required that the Ricci scalar is finite for solutions to \eq{deSitter}. If no such solutions were to be found, the action would take its extremal values at the boundaries of the definition domain, $R=0$, and $1/|R|=0$, without necessarily obeying \eq{deSitter}.

\subsection{Real de Sitter solutions}

\begin{figure*}[t]
\begin{center}
\unitlength0.001\hsize
\begin{picture}(1000,500)
\put(100,-10){\includegraphics[width=.75\hsize]{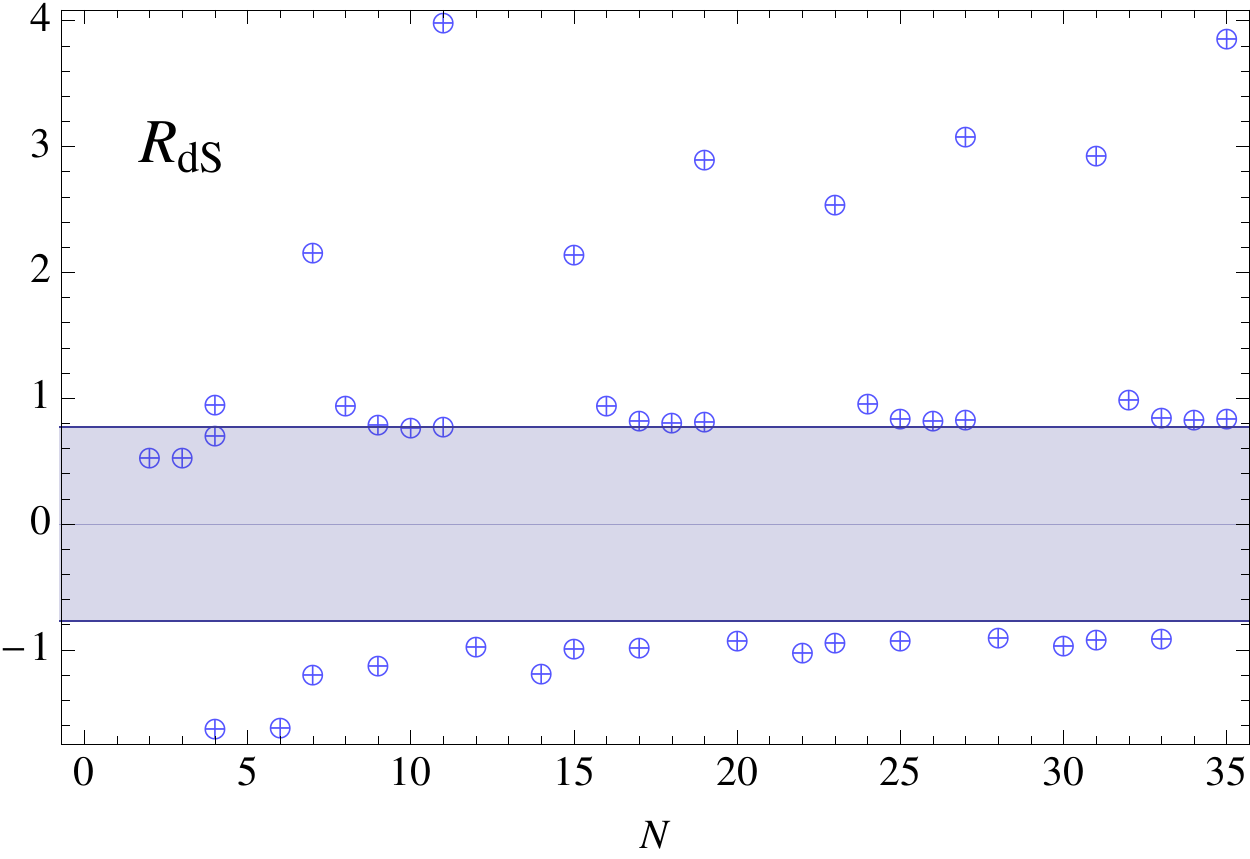}}
\end{picture}
\caption{\label{deSitterRadius} Shown are all real de Sitter and anti-de Sitter solutions for each polynomial approximation order $N$ (crossed circles). Numerical values are given in Tab.~\ref{tdS}. The shaded area indicates the estimated radius of convergence \eq{deSitterLB}. The eight-fold periodicity pattern in the signs of couplings leads to solutions close to the radius of convergence. We conclude the absence of real de Sitter or anti-de Sitter solutions for small curvature scalar within the radius of convergence of the polynomial expansion.}
\end{center}
\end{figure*}

We can look for solutions to \eq{deSitter} at each order $N$ in the approximation by plotting the LHS of the equation and looking for stable zeros for all approximation orders $N$. Prior to this, we need to estimate the radius of convergence for $E(R)$ accurately, to ensure solutions are in the physical domain. To that end, we expand the stationarity condition \eq{deSitter}  as a polynomial in the Ricci scalar, 
\beq\label{zeta}
E(\R)=\sum_n\xi_n R^n\,,
\eeq 
and search for a fit for  the expansion coefficients $\xi_n$ as
\begin{equation}\label{desittercos}
\xi_n= A\,\frac{\cos(n\,\phi+ \Delta)}{(R_c)^n} \,.
\end{equation}
This provides us with an estimate for the radius of convergence for $E(R)$. Fitting the four free parameters in \eq{desittercos} to the data, we find
\begin{equation}\label{fitdeSitter}
\begin{array}{rl}
A&=0.5282\\[.5ex]
R_c&=0.8584\\[.5ex]
\phi&=0.8226\\[.5ex]
\Delta&=2.1334
\end{array}
\end{equation}
for the amplitude, the radius, the angle and the shift, respectively. Fig.~\ref{pdeSitterCos} shows that \eq{fitdeSitter} provides a good fit to the data. The angle is similar to the one found in \eq{fit1} but the radius is slightly smaller. 
We additionally adopt the generalised ratio test. Since $R^2$ is a `zero-mode' of \eq{deSitter} there will be no terms proportional to $\R^2$ in \eq{zeta}. Therefore we take $n \geq 3$ when determining $\R_c(m)\equiv\min_n[\R_{c,m}(n)]$, and average m over values $8 \le m \le 31$, see Sect.~\ref{sing}. Using this method we obtain the lower bound estimate
\begin{equation}\label{deSitterLB}
\RLB \approx 0.77 \pm 5\% 
\end{equation}
which is lower than the corresponding lower-bound estimate obtained 
from the polynomial expansion of $f(\R)$. The reason for this is that the equation of motion contains a derivative of $f(\R)$ which is  more sensitive to the polynomial  approximation, lowering the expected radius of convergence. A summary of the various radii of convergence including error estimates is given in Tab.~\ref{tRadius}. 

With these results at hand we now return to the search for zeros of \eq{zeta}. In fact, real solutions to \eq{deSitter} can be found at some orders in the approximation. 
In Fig.~\ref{deSitterRadius}, we display all real de Sitter solutions for each and every approximation order $N$, indicated by crossed circles, and the estimate for the radius of convergence \eq{deSitterLB}, indicated by the shaded area. Numerical results are shown in Tab.~\ref{tdS}. These solutions may be considered as physical provided they arise
within the radius of convergence of the expansion and persist to high orders in the expansion. Some solutions accumulate close to $\Rds\approx 0.8-0.9$ and $\Rds\approx -(0.9-1.0)$, and appear to be at the boundary of the radius of convergence  \eq{root} and \eq{RL} as determined from $f(R)$. Most notably, we find that de Sitter solutions  occur within the radius of convergence, as determined from $E(R)$, but only low orders in the approximation, without persisting to higher orders. We conclude that the data does not offer evidence for a real de Sitter solution at small Ricci curvature.

\begin{center}
\begin{table}
\addtolength{\tabcolsep}{1pt}
\setlength{\extrarowheight}{1pt}
\scalebox{0.95}{
\begin{tabular}{c|G|c|G|c|G}\hline\hline
\rowcolor{LightGreen}
\ \bf approx.\ & \ \ \bf positive root\ \  &\ \ \bf negative root\ \ &\multicolumn{3}{c}{\cellcolor{LightGreen}\bf complex root}\\ \hline
\rowcolor{LightYellow}
${}\quad N\quad$& $\quad\Rds>0\quad$&$\quad\Rds<0\quad$&$\quad{\rm Re}\,\Rds\quad$&$\quad{\rm Im}\,\Rds\quad$&$\quad|\Rds|\quad$\\ \hline\hline
  2 & 0.5171& &&&\\
 3 & 0.5174 &&&&\\
 4 & 0.6949& $-$1.637 &&&\\
 5&&& 0.5630 & 0.2095 & 0.6008\\
6 & &$-$1.626& 0.5459 & 0.2023 & 0.5822  \\
 7 & 2.148 & $-$1.206& 0.5421 & 0.2089 & 0.5810\\
 8 & 0.9338 && 0.5371 & 0.2262 & 0.5827\\
 9 & 0.7856 & $-$1.135 & 0.5428 & 0.2345 & 0.5913 \\
 10 & 0.7580&& 0.5531 & 0.2363 & 0.6015 \\
 11 & 0.7692 && 0.5547 & 0.2357 & 0.6027 \\
 12 && $-$0.981& 0.5596 & 0.2420 & 0.6097 \\
 13&&&0.5592 & 0.2452 & 0.6106\\
 14 && $-$1.196& 0.5610 & 0.2460 & 0.6125 \\
 15 &2.130& $-$0.997& 0.5590 & 0.2453 & 0.6105 \\
 16 & 0.9336&& 0.5592 & 0.2436 & 0.6099 \\
 17 & 0.8158 & $-$0.991& 0.5626 & 0.2420 & 0.6124\\
 18 & 0.7986 && 0.5652 & 0.2415 & 0.6147\\
 19 & 0.8079& & 0.5648 & 0.2414 & 0.6142\\
 20& & $-$0.931 & 0.5645 & 0.2425 & 0.6144\\
 21&&& 0.5635 & 0.2420 & 0.6133\\
 22 && $-$1.029 & 0.5642 & 0.2417 & 0.6138 \\
 23 &2.528 & $-$0.947 & 0.5635 & 0.2418 & 0.6132\\
 24 & 0.9514&& 0.5631 & 0.2418 & 0.6128 \\
 25 & 0.8286 & $-$0.938 & 0.5638 & 0.2415 & 0.6134\\
 26 & 0.8141 && 0.5648 & 0.2413 & 0.6142\\
 27 & 0.8202 && 0.5646 & 0.2413 & 0.6140\\
28 && $-$0.910 & 0.5647 & 0.2420 & 0.6143 \\
 29 && & 0.5644 & 0.2419 & 0.6141 \\
 30 && $-$0.970 & 0.5648 & 0.2418 & 0.6144\\
 31 & 2.923& $-$0.924 & 0.5645 & 0.2418 & 0.6141\\
 32 & 0.9824 && 0.5643 & 0.2417 & 0.6139\\
 33 & 0.8365 & & 0.5647 & 0.2416 & 0.6142 \\
 34 & 0.8235 && 0.5653 & 0.2414 & 0.6147 \\
 35 & 0.8270& & 0.5651 & 0.2414 & 0.6145\\ \hline\hline
\rowcolor{LightYellow}
\bf mean (all)&n/a&n/a&0.5560& 0.2377& 0.6084\\ \hline\hline
\rowcolor{LightGreen}
\bf mean (cycle)&&&0.5647& 0.2417& 0.6143\\
\rowcolor{LightGreen}
\bf st.~dev.~(\%)&&&$\pm0.06\%$& $\pm0.08\%$& $\pm0.04\%$\\ \hline\hline
\end{tabular}
}
\caption{\label{tdS} Selected de Sitter solutions from the polynomial approximation up to including $N=35$. For each order the smallest positive, the smallest negative, and the complex root with smallest length are shown, provided they exist. The mean values taken over all data and a cycle of the eight highest orders, are also given. The standard deviation is very small, establishing that the complex conjugate pair of de Sitter solutions is very stable and converges rapidly. The  positive roots for  $N=2$ to $11$   agree with \cite{Bonanno:2010bt}. Notice that the real roots display an eight-fold periodicity pattern. }
\end{table} 
\end{center}

\begin{center}
\begin{table}[t]
\addtolength{\tabcolsep}{1pt}
\setlength{\extrarowheight}{1pt}
\begin{tabular}{Gc|G|l}\hline\hline
\rowcolor{LightGreen}
\rowcolor{LightYellow}
${}^{}$\quad\quad\quad&\ \ approximation \ \ &  \quad radius estimate \quad &
\cellcolor{LightYellow}
\quad\quad\quad \quad\quad\quad\quad\quad info \ \ 
\\ \hline
 a)& $N=11$ & $0.99$& eff~action, \eq{N11} -- from Ref.~\cite{Bonanno:2010bt}  \\
b)& $N=11$ & $1.00\pm 20\%$& eff~action, root test \eq{N11new} \\ 
c)& $N=35$ & $0.91\pm 5\%$ & eff~action, fit and root tests, \eq{fit1}, \eq{root} \\
d)& $N=35$ & $0.82\pm 5\%$& eff~action, lower bound, \eq{RL} \\ \hline
e)& $N=35$ & $0.86$& eq of motion, fit and root tests, \eq{fitdeSitter}\\
f)& $N=35$ & $0.77\pm 5\%$& eq of motion, lower bound, \eq{deSitterLB} \\ 
\hline\hline
\end{tabular}
\caption{\label{tRadius}  Summary of  estimates for the radius of convergence based on
the action and the equations of motion, also comparing different approximation orders. Errors are statistical, roughly a standard deviation.}
\end{table} 
\end{center}

\subsection{Comparison}

At this point it is useful to make contact with the results of \cite{Bonanno:2010bt}. While our results up to order $N=11$ agree with \cite{Bonanno:2010bt}, we are reaching the exact opposite conclusion. The reason for this is a subtle one, related to the slow convergence of  the polynomial expansion. The eight-fold periodicity pattern in the couplings, \eq{8}, entails de Sitter solutions around $\Rds \approx 0.77$ for  nine of the first eleven approximation orders considered in \cite{Bonanno:2010bt},  see Tab.~\ref{tdS}, suggesting that these are viable. The underlying eightfold periodicity only became visible starting from approximation order $N=11-14$ onwards, based on the sign pattern of couplings (Tab.~\ref{tFP}) or their magnitude (Fig.~\ref{pCos}), respectively. Then, for all $N>11-14$, a de Sitter candidate arises in four out of eight cases. However, with increasing $N$, the de Sitter candidates no longer reside within the domain of validity. This follows from the various estimates for the radii of convergence summarised in Tab.~\ref{tRadius}. Firstly, using the polynomial approximation for $f(R)$, the uncertainty in the radius of convergence at order $N=11$ comes out large, about $20\%$, Tab.~\ref{tRadius} $b)$. This is going down to the 5\% range at $N=35$, Tab.~\ref{tRadius} $c)$ and Tab.~\ref{tRadius} $d)$. The positive de Sitter solutions of \cite{Bonanno:2010bt} sit close to the boundary of this error estimate. 
Secondly, once the polynomial approximation for $E(R)$ is analysed to estimate  the domain of validity, the radius comes out smaller by about 5\%, see Tab.~\ref{tRadius} $e)$ and Tab.~\ref{tRadius} $f)$. Combining the decrease of the radius of convergence and its error bar  (with increasing $N$) with the slight growth of real de Sitter solutions $R_{\rm dS}>0$ (see Tab.\ref{tdS}), it  becomes evident that these solutions slip out of the domain of validity.

\subsection{Complex de Sitter solutions}
\begin{figure*}[t]
\begin{center}
\unitlength0.001\hsize
\begin{picture}(1000,660)
\put(140,-20){\includegraphics[width=.7\hsize]{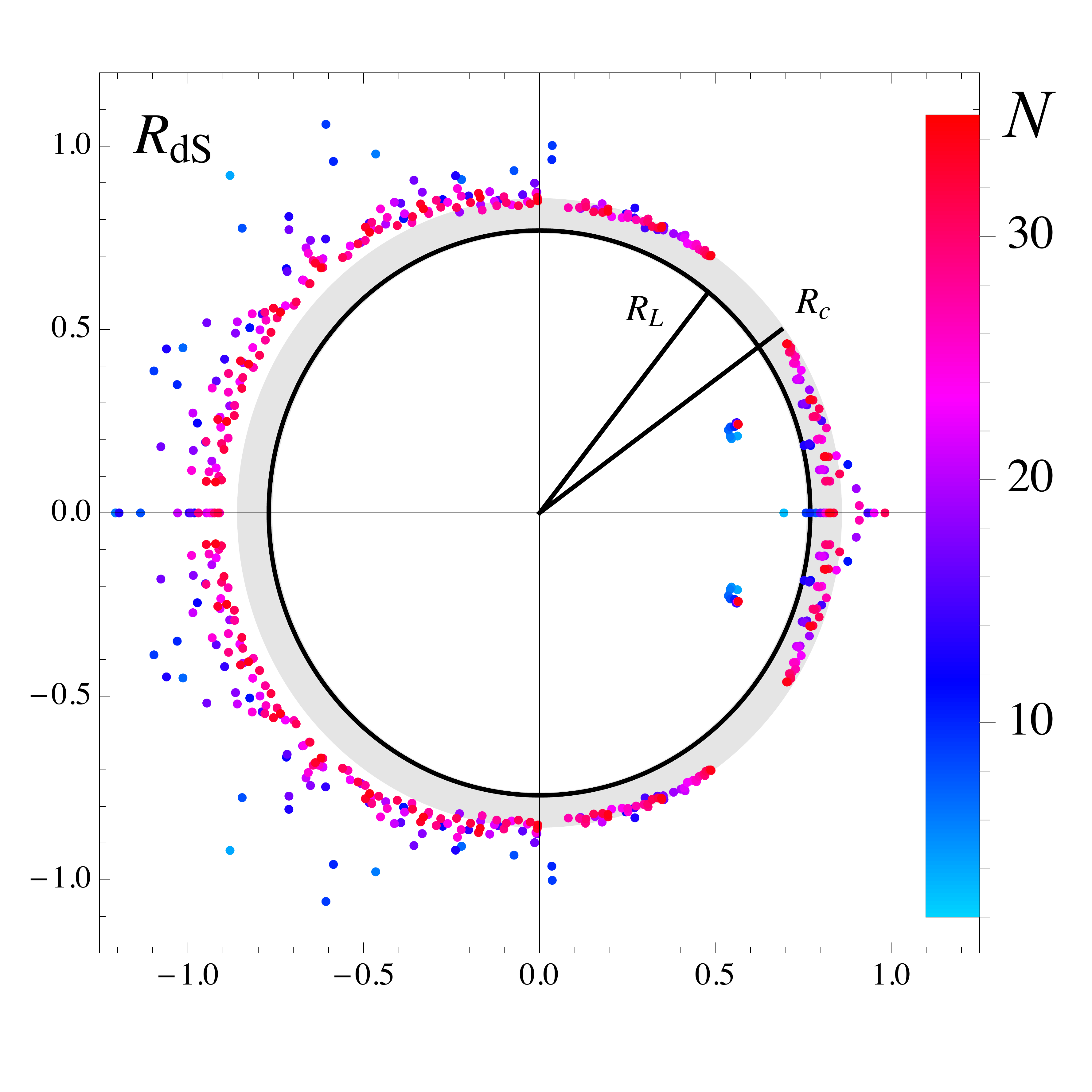}}
\end{picture}\caption{\label{deSitterComplex} Real and complex de Sitter solutions $\Rds$ in the plane of complexified Ricci scalar in units of $k^2$, colour-coded from blue to red with increasing approximation order $N$.  Higher order data points are shown on top of the lower order ones. The lower bound $\RLB$ for the radius of convergence,  \eq{deSitterLB}, is indicated by a black circle. The gray ring covers the range between $\RLB$ and $R_c$, see \eq{fitdeSitter}. Most roots with positive real part accumulate close to the radius $R_c$.
Notice the existence of a unique and stable de Sitter solution with a positive real and a small imaginary part   within the radius of convergence  for all approximation orders $N\ge 5$. }
\end{center}
\end{figure*}

For completeness, we check whether de Sitter solutions exist within the radius of convergence for complex Ricci scalar. Within the polynomial approximation to order $N$, the de Sitter condition \eq{deSitter} will naturally have up to $N-1$ different solutions in the complex plane. In Fig.~\ref{deSitterComplex}, we display all of these solutions for all orders $N$. We find that most solutions are outside the estimated radius of convergence. Those with positive real part accumulate close to the distance $R_c$ \eq{fitdeSitter}. Interestingly, within the radius of convergence \eq{deSitterLB}, there exist exactly two solutions with $|\Rds|<R_L$, a complex conjugate pair, for each approximation order as soon as $N\ge 5$. The complex de Sitter solution has a positive real and a small imaginary part,
\begin{equation}\label{RdS}
\begin{array}{rl}
{\rm Re}\,\Rds&=\ \ \,0.5647\pm0.02\%\\[.5ex]
{\rm Im}\,\Rds&=\pm0.2417\pm0.02\%\\[.5ex]
|\Rds|&=\ \ \, 0.6143\pm0.02\%\,.
\end{array}
\end{equation}
Notice that the solution is particularly stable from order to order, reflected by the tiny standard deviation. 
Quantitatively, our results are summarized in Tab.~\ref{tdS}, where the smallest positive, negative, and the smallest complex solution per order are shown. 
  
\begin{figure*}[t]
\begin{center}
\unitlength0.001\hsize
\begin{picture}(1000,320)
\put(10,0){\includegraphics[width=.485\hsize]{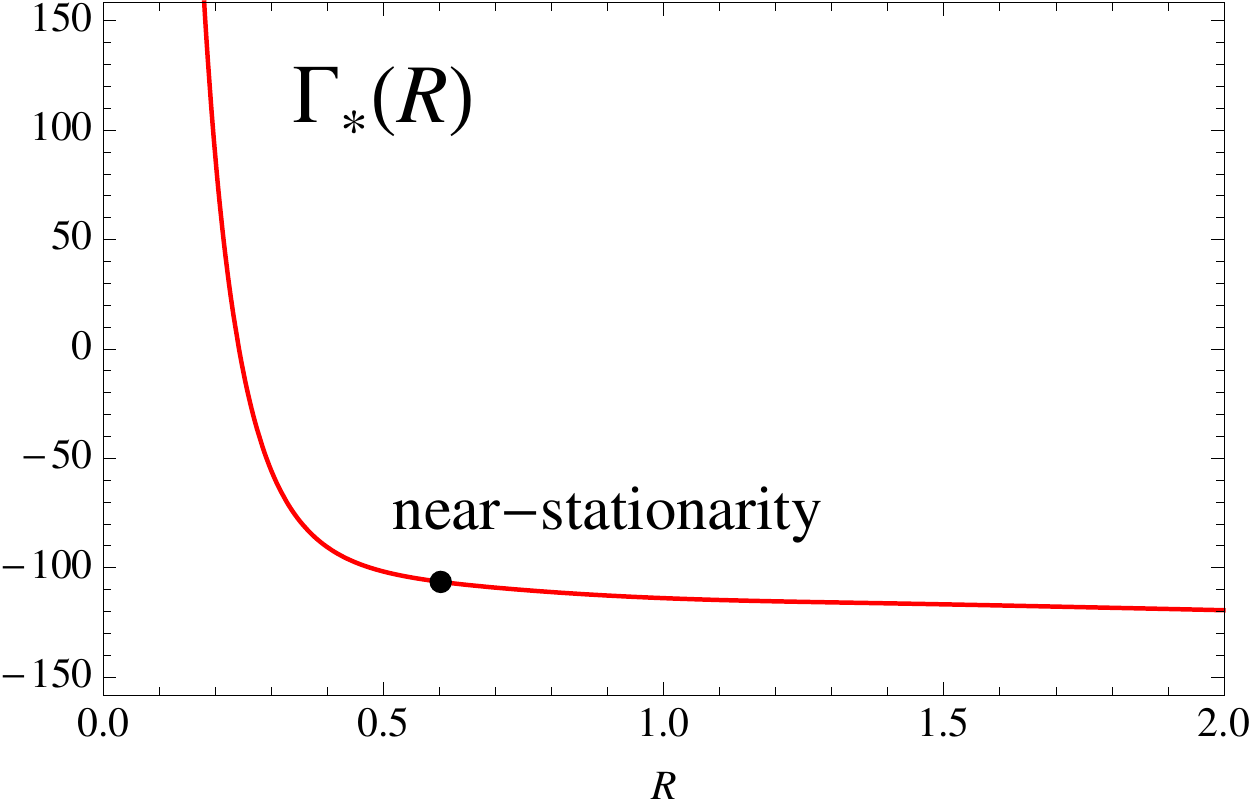}}
\put(510,0){\includegraphics[width=.46\hsize]{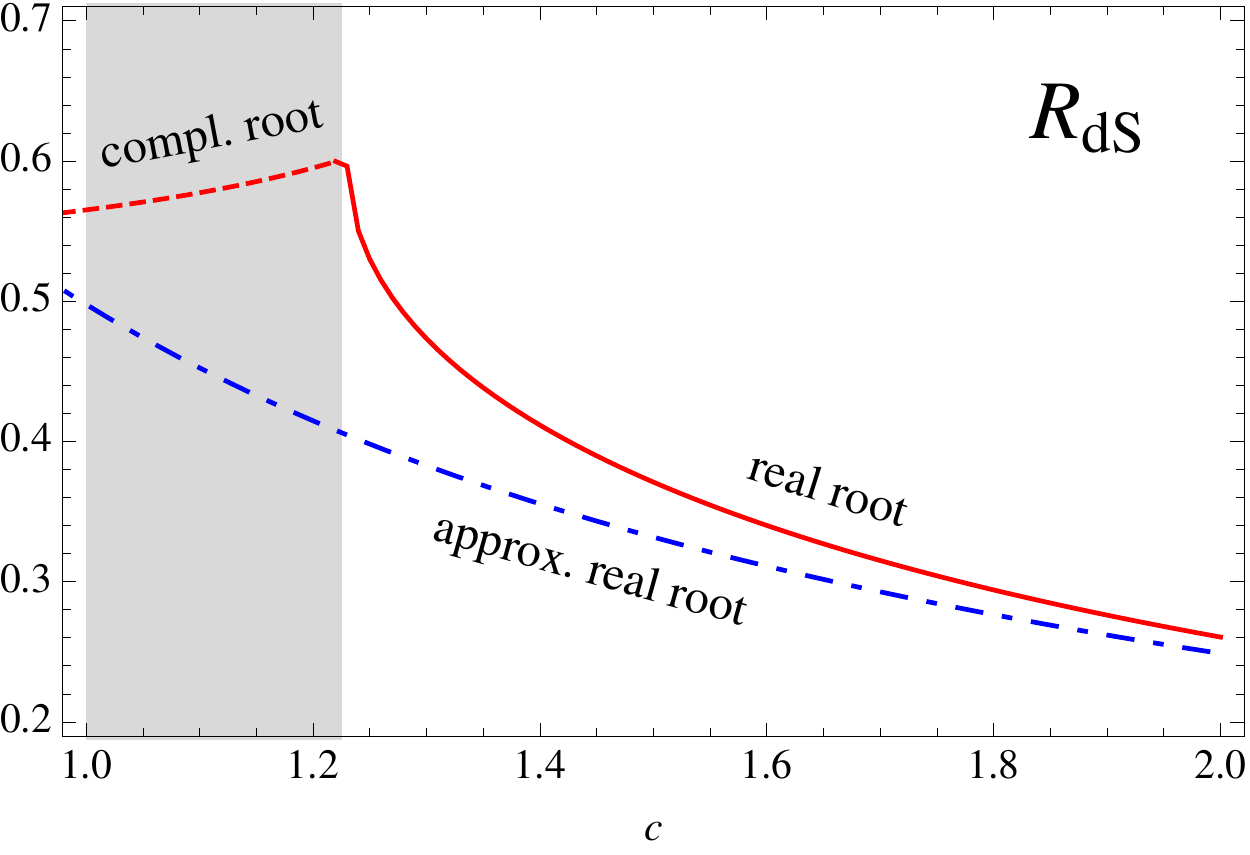}}
\end{picture}
\caption{\label{pRdSc} Left panel: the fixed point effective action \eq{Gammak} as a function of scalar curvature. The near-stationary point \eq{Rnear} is indicated by a full dot. Right panel: solutions $R_{\rm dS}$ to the stationarity condition assuming that  $\lambda_*$ at the UV fixed point is reduced by a fudge factor $c>1$, while $g_*$ and all other fixed point couplings remain unchanged. The complex de Sitter solution (dashed red line) then becomes a  real one (full red line) as soon as $c\simeq 1.25$. Once a real solution  exists, its location is well-approximated by  \eq{dSapprox}
indicated by the blue dashed-dotted line. All higher order couplings only add small corrections.}
\end{center}
\end{figure*}

\subsection{From near-stationary to stationary points}\label{near}
We have established that the fixed point theory does not offer a stationary point with a real de Sitter solution, at least  in the regime for small scalar curvature. On the other hand, a near-stationary solution is located close to where $\partial\Gamma/\partial R$ is smallest. In fact, solving $\min_R(2f-Rf')$ for $R$ one finds the near-stationary solution
\beq\label{Rnear}
R_{\rm ns}\approx 0.602\,,
\eeq
whose location  is dictated by the nearby
exact complex root \eq{RdS} with $R_{\rm ns}\approx |\Rds|$. For illustration, the left panel of Fig.~\ref{pRdSc} displays the near-stationary solution, showing that it corresponds to a nearly flat effective fixed point action $\Gamma_*(R)$ starting from curvatures $R$ around and above  \eq{Rnear}. We may conclude that the near-stationarity leads to near-de Sitter behaviour in the deep UV for scalar curvatures around \eq{Rnear}.

Up to now we have discussed a model of quantum gravity with high-order interactions in the Ricci scalar. In a more extended study one should retain other curvature invariants as well. If the UV fixed point is a feature of the full theory, we expect that the impact of further curvature invariants is to change the fixed point couplings. 
To mimick the influence of (neglected) curvature invariants, and to check how this is influencing the existence (or not) of de Sitter solutions, we have introduced fudge factors by varying the values of the first few couplings $\lambda_0$, $\lambda_1$, $\lambda_3$ and $\lambda_4$ by $\pm 50\%$. (Notice that de Sitter solutions are insensitive to the coupling $\lambda_2$). One finds that variations of  $\lambda_3$ and $\lambda_4$ do barely change the nature of the de Sitter solution. Instead, varying $\lambda_0$ and $\lambda_1$ has a more substantial influence. For example, reducing $\lambda_1$ by a fudge factor of $c>1$ without changing $g_*$ nor the other couplings, leads to a real de Sitter solution once $c\simeq 1.25$, see  Fig.~\ref{pRdSc} (left panel).  This can also be understood from Fig.~\ref{pdeSitter}: lowering $\lambda_0$ lowers the entire curve without changing its shape, which can lead to real de Sitter solutions. We conclude that stationary solutions may exist close to the near-stationary 
solution  of the fixed point theory described by \eq{expansion0}.

Interestingly, if a real de Sitter solution exists, it is already well-approximated by the leading order estimate
\beq\label{dSapprox}
R_{\rm dS}\simeq -2\,\frac{\lambda_0}{\lambda_1}\equiv 4\lambda_*\,.
\eeq
If the root is complex, then \eq{dSapprox} offers a fair approximation to its real part. In either case, the impact of higher order couplings on the numerical value for $R_{\rm dS}$ appears to be small. The right panel of Fig.~\ref{pRdSc}  compares the approximate solution \eq{dSapprox} (blue dashed-dotted line) with the full, real or complex, solution (full or dashed red lines, respectively).  We conclude that if a real de Sitter points exists, its value is fixed primarily by the dimensionless cosmological constant $\lambda_*$.

\subsection{De Sitter solutions at large curvature}
\begin{figure*}[t]
\begin{center}
\unitlength0.001\hsize
\begin{picture}(1000,450)
\put(140,0){\includegraphics[width=.7\hsize]{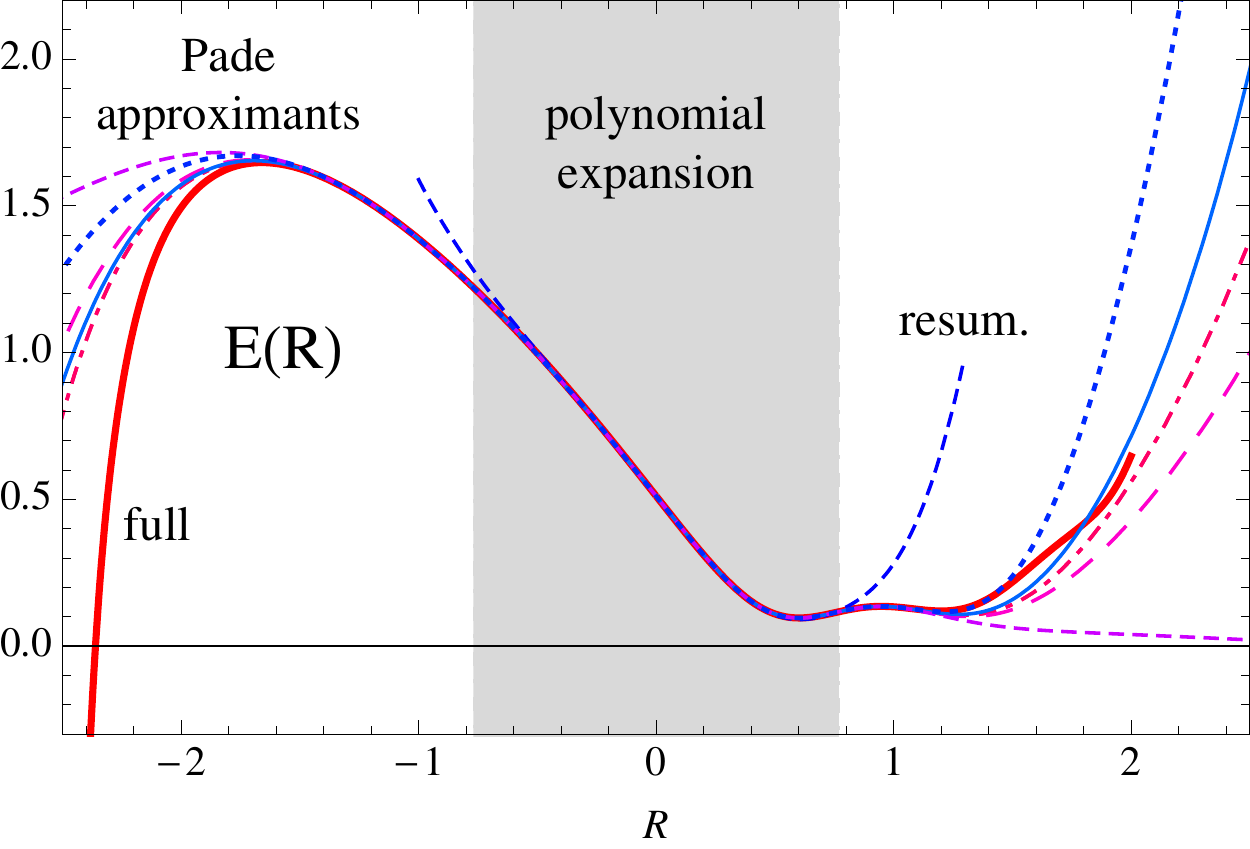}}
\end{picture}
\caption{\label{pdeSitter}Shown is the function $E(R)=2f(R)-R\,f'(R)$ at the fixed point obtained by the full numerical integration of \eq{df''} (thick red line) in comparison with several Pad\'e approximants thereof  ([20/16]: dotted line, [17/14]: dashed-dotted, [14/12]: long-dashed, [16/15]: thin  full, [16/17]: short dashed), and the resummation \eq{resum}, \eq{resum2} (resum). The shaded area indicates the radius of convergence \eq{deSitterLB} of the underlying polynomial approximation \eq{zeta}.
Stationary points of the fixed point action correspond to zeros of $E(R)$. Despite of differences in their asymptotic behaviour, the Pad\'e approximants agree very well with the numerical solution within  $ -1.5\lta R\lta 1.2$, much beyond the range $R_L$ \eq{deSitterLB} (see main text).}
\end{center}
\end{figure*}

Further real de Sitter solutions may exist for larger Ricci curvature and outside the radius $\R_L,$ but within the region where a full numerical or resummed result for $E(R)$  exist. To investigate these cases, we analyse the equation of motion $E(R)$ using numerical integration in the regime $-2.5 \leq \R \leq 2$, together with Pad\'e approximants, of various kinds, for the polynomial expansion of $E(R)$. 
Our numerical integration becomes unreliable close to $R\approx 2.006$ and close to $R\approx -2.541$,  which prevents the solution being continued to larger values of $|R|$. 
The numerical solution does not show any de Sitter solutions satisfying the equation of motion within the region $-2 \leq \R \leq 2$.  However, the function $E(R)$ becomes very small in the regime $R\approx 1$, see Fig.~\ref{pdeSitter}. To confirm our result, we additionally approximate $E(R)$ through  Pad\'e approximants as we did in Sec.~\ref{pade} for $f(R)$. Specifically,  in Fig.~\ref{pdeSitter}, we display the Pad\'e approximants [20/16] (dotted line), [17/14] (dashed-dotted), [14/12] (long-dashed), [16/15] (thin  full), and [16/17] (short dashed), together with the numerical integration (full red line). The shaded area indicates the radius of convergence of the underlying polynomial approximation, \eq{deSitterLB}.

We find very good agreement between Pad\'e approximants and the numerical solution beyond the radius of convergence $R_L$, up to the range $-1.5\lta R\lta 1.2$, and in the regime where $E(R)$ becomes small numerically. Once $1.2 \lta R$, numerical differences arise. The equation of motion $E(R)$ is more sensitive to the large-$R$ behaviour of the Pad\'e approximants than the fixed point action $f(R)$. The numerical result agrees best with Pad\'e approximants whose asymptotic behaviour is given by $\propto R^2$ or $\propto R^3$.  In all cases, no de Sitter solution shows up for $R\lta 2$. Within the uncertainties of our study we cannot exclude de Sitter solutions for even larger values of the curvature.  

For negative scalar curvature $R\lta -1.5$, we also find that differences between Pad\'e approximants become significant. An anti-de Sitter solution is found  from the numerical solution at
\begin{equation}\label{AdS}
R_{\rm AdS} \approx - 2.36138\,.
\end{equation} 
This result should be cross-checked with Pad\'e approximants, as these allow extrapolation to the regime where the numerical integration is no longer applicable.  The Pad\'e approximants also predict a de Sitter solution, roughly in the range $R_{\rm AdS} \approx - (2.5 - 3.5)$, although the location is clearly not determined accurately. We may conjecture that an anti-de Sitter solution at scalar curvature $|R|$ of the order of a few exists  within the present approximation.

\subsection{Discussion}
We have established the absence of real de Sitter solutions in the fixed point regime for small Ricci scalar $|R|<R_L$, based on high orders in the polynomial approximation for the fixed point. 
In addition, we have also found evidence for the absence of de Sitter solutions within the wider range $-2\lta R\lta 1.5$. Here, resummation techniques and numerical integration have been used to establish the result.  We have not explored the possibility of de Sitter solutions for large curvature scalar beyond the domain of validity of our study, but it is conceivable that a de Sitter solution for finite $R$ exists. The anti-de Sitter solution close to  \eq{AdS} gives a first indiction of this. 
Further work is needed to obtain flow equations valid for all values of $R$ where de Sitter solutions may be found. 
In \cite{Dietz:2012ic} it has been shown, albeit in a different RG scheme, that a valid fixed point solution may display a power-law behaviour asymptotically, modulo oscillating logarithmic corrections. The existence of de Sitter solutions then depends on the coefficients of this asymptotic behaviour. Similar results may apply to our model and it remains to be seen whether or not they would offer de 
Sitter solutions \eq{deSitter} for asymptotically large $R\gg 1$. 
If no solutions to the de Sitter condition \eq{deSitter} for finite $R$ can be found, the action takes its extremal values at the boundaries of vanishing or infinite curvature. In the limit $|R|\to 0$, the action becomes maximal because of $f(R=0)>0$, suggesting that a global minima is achieved either for large and finite $R$ outside the range considered here, 
or at asymptotically large $|R|=\infty$.

The ultraviolet fixed point  displays a stable complex de Sitter  solution with positive real part, close to the real axis, for each and every approximation order starting at $N=5$; see \eq{RdS}. The relevancy of this solution for cosmological applications is unclear: it would seem to imply that 
a ``near de Sitter" phase with inflationary expansion in the fixed point regime of $f(R)$ gravity could exist, as long as the imaginary part is sufficiently small (as it is here).  Small variations in the fixed point couplings 
show that the complex de Sitter solution becomes real once other curvature invariants or matter are taken into account (Fig.~\ref{pRdSc}).  A first  example for this has been given in \cite{Falls:2017lst}. There, it has been established that Ricci tensor fluctuations such as in quantum gravity  with action $\propto  F(R^2)+R\cdot Z(R^{\mu\nu}R_{\mu\nu})$   induce de Sitter solutions even at an asymptotically safe UV fixed point.

\section{\bf Conclusions}\label{C}

We have studied the availability of inflationary solutions for cosmology from  asymptotically safe models of quantum gravity in the high-energy regime.
The relevant information -- encoded in the full fixed point effective action -- requires a good understanding of the fixed point couplings themselves, including the relevant as well as the higher order irrelevant interactions. 
The reason for this relates to a  singularity (cut or pole) in the plane of complexified Ricci curvature whose fingerprint is the characteristic  convergence pattern of gravitational couplings in the physical domain (Fig.~\ref{Poles}). This is reminiscent of other quantum field theories at a critical point \cite{Litim:2016hlb,Juttner:2017cpr},
and required high polynomial orders to determine couplings with good accuracy including errors. 
Our findings then exclude the existence of de Sitter solutions in the deep UV regime for small Ricci curvature up to the order of the RG scale $R\propto k^2$.  Interestingly though, the theory does display a complex de Sitter solution with small imaginary part, corresponding to near de Sitter behaviour in the physical regime (Fig.~\ref{deSitterComplex}).  Our findings also serve as a word of caution:  while low-order approximations of the gravitational effective action often detect the UV fixed point correctly, e.g.~\cite{Souma:1999at,
Lauscher:2001ya,Lauscher:2002sq,Litim:2003vp,Fischer:2006fz,Fischer:2006at,Litim:2006dx,Codello:2006in,
Codello:2007bd,Machado:2007ea,Codello:2008vh,Litim:2008tt}, the existence or not of de Sitter solutions in the scaling regime is much more sensitive to the precise strength of higher derivative interactions \cite{Falls:2013bv,Falls:2014tra}. Here, low order approximations do not prove sufficient  to determine de Sitter solutions reliably (Tab.~\ref{tdS}).

Under mild variations of fixed point couplings, in particular the cosmological constant, we  have established that near de Sitter solutions  bifurcate into real ones, showing that the purely gravitational fixed point is located at the boundary of theories with or without small-field de Sitter solutions (Fig.~\ref{pRdSc}). For this reason,  the inclusion of further curvature invariants such as Ricci or Riemann tensor invariants \cite{Falls:2017lst}, derivatives thereof, or the use of improved approximation schemes  \cite{Litim:1998nf,Litim:2002ce,Litim:2002hj}, can tip the balance from near- to real de Sitter solutions. Recent work suggests that the presence of Ricci tensor interactions  $\sim (R_{\mu\nu}R^{\mu\nu})^n$ is of particular relevance  for this to happen \cite{Falls:2017lst}.  In a different vein, it will  be equally important to understand how gravitational $f(R)$ actions behave away from the fixed point, and whether inflationary phases of cosmology can be achieved along  trajectories which connect the UV fixed point with the IR regime of classical general relativity \cite{Hindmarsh:2011hx,Hindmarsh:2012rc}.  
Overall, it then is  conceivable that asymptotically safe quantum gravity can trigger a phase of inflationary expansion in the very early universe, but more work is required to settle the specifics for this.

On the  technical side, a novelty of our study is the use of resummation techniques such as Pad\'e  to obtain improved results for the gravitational effective action.  Intriguingly,
Pad\'e approximants ameliorate the result in field space far beyond the polynomial radius of convergence  (Figs.~\ref{pResum},~\ref{pResumPade} and \ref{pdeSitter}). It is conceivable that the technique will prove useful for other theories as well. 
 Together with infinite order approximations, resummations have enlarged the domain of validity of our study. It is noteworthy that the resummed results are in agreement with the full numerical solutions even beyond the radius of convergence, showing that  the set of polynomial couplings encode more physics information than previously exploited. On the other hand, the large-curvature asymptotics -- which serves as an input into Pad\'e approximants -- is not well predicted within the present setup. The reason for this is that the underlying heat kernel expansion becomes less reliable for large curvature. We hope to revisit these aspects in due course.

\section*{\bf  A\lowercase{cknowledgements}}
We thank Mark Hindmarsh and Edouard Marchais for discussions. This work is supported by the Science Technology and Facilities Council (STFC) under grant number ST/G000573/1, and by the A.S. Onassis Public Benefit Foundation grant F-ZG066/2010-2011 (KN). DL acknowledges support by the National Science Foundation under Grant No.~PHYS-1066293, and  hospitality of the Aspen Center for Physics.

\bibliography{deSitterFinal.bib}
\bibliographystyle{JHEP}

\end{document}